\documentclass[5p, number, sort, compress]{elsarticle}

\usepackage{color}
\usepackage{amsmath} 
\usepackage{amssymb}
\usepackage{graphicx}
\usepackage[caption=false]{subfig} 
\usepackage{dcolumn}
\usepackage{bm}
\usepackage{hyperref}
\usepackage{verbatim}

\newcommand{\phicr}{\phi_\mathrm{cr}}

\newcommand{\Tscr}{T^*_\mathrm{cr}}

\newcommand{\GM}{\hat{G}_\mathrm{M}}

\newcommand{\GMM}{\hat{G}_\mathrm{MM}}
\newcommand{\GMI}{\hat{G}_\mathrm{MI}}
\newcommand{\GIM}{\hat{G}_\mathrm{IM}}
\newcommand{\GII}{\hat{G}_\mathrm{II}}

\newcommand{\fel}{f_\text{el}}
\newcommand{\kr}{\tilde{k}}
\newcommand{\chiel}{\chi_{\rm el}}

\newcommand{\expkR}{\zeta}

\newcommand{\Tscrp}[1]{T^{* #1}_\mathrm{cr}}

\begin{document}

\title{Random-phase-approximation theory for sequence-dependent, biologically
functional liquid-liquid phase separation of intrinsically disordered proteins}

\author[biochem,sickkids]{Yi-Hsuan Lin}

\author[biochem,molgene]{Jianhui Song\tnoteref{present_add}}

\author[sickkids,biochem]{Julie D. Forman-Kay}

\author[biochem,molgene]{Hue Sun Chan\corref{cor1}}\ead{chan@arrhenius.med.toronto.edu}

\address[biochem]{Department of Biochemistry, University of Toronto, 1 King's College Circle, Toronto, Ontario M5S 1A8, Canada}
\address[sickkids]{Molecular Structure and Function Program, Hospital for Sick Children, 686 Bay Street, Toronto, ON M5G 0A4, Canada}
\address[molgene]{Department of Molecular Genetics, University of Toronto, Toronto, 1 King's College Circle, Ontario M5S 1A8, Canada}

\cortext[cor1]{Corresponding author}
\tnotetext[present_add]{Present address: School of Polymer Science and Engineering, Qingdao University of Science and Technology, Shandong, China.}

\date{2016/09/22}	

\begin{abstract}

Intrinsically disordered proteins (IDPs)
are typically low in nonpolar/hydrophobic 
but relatively high in polar, charged, and aromatic amino acid compositions. 
Some IDPs undergo liquid-liquid phase separation in the aqueous milieu of 
the living cell. The resulting phase with enhanced IDP concentration 
can function as a major component of membraneless organelles that, by 
creating their own IDP-rich microenvironments, stimulate critical biological 
functions. IDP phase behaviors are governed by their amino acid sequences. 
To make progress in understanding this sequence-phase relationship, 
we report further advances in a recently introduced application of
random-phase-approximation (RPA) heteropolymer theory to account for 
sequence-specific electrostatics in IDP phase separation. Here we examine
computed variations in phase behavior with respect to block length and 
charge density of model polyampholytes of alternating equal-length charge 
blocks to gain insight into trends observed in IDP phase separation. As a
real-life example, the theory is applied to rationalize/predict
binodal and spinodal phase behaviors of the 236-residue N-terminal disordered
region of RNA helicase Ddx4 and its charge-scrambled mutant for 
which experimental 
data are available. Fundamental differences are noted between the 
phase diagrams predicted by RPA and those predicted by mean-field 
Flory-Huggins and Overbeek-Voorn/Debye-H\"uckel theories. In the RPA
context, a physically plausible dependence of relative permittivity 
on protein concentration can produce a cooperative effect in favor 
of IDP-IDP attraction and thus a significant increased tendency to 
phase separate. Ramifications of these findings for future development 
of IDP phase separation theory are discussed.

\end{abstract}

\maketitle

\section{Introduction}


Globular proteins have been the predominant focus of structural 
biology since the folded myoglobin structure was revealed 
by X-ray crystallography in 1958 \cite{Kendrew58}. However, while
many proteins need to fold to a relatively fixed structure for 
biological activity, it is now clear --- after more than one and a half decade 
of intense research~\cite{Uversky00,Dunker01,Tompa02,Nussinov2003,Dyson2005}
--- that intrinsically disordered proteins (IDPs) perform key functions
in cellular processes~\cite{tompa12,julie12,FormanKay13,vanderLee14,Liu2014,Chen15,PappuCOSB,Wright2015,Veronika2016,Fuxreiter2016}.
This advance led to the recognition that different biological 
functions can be bestowed upon protein conformations of various degrees of
flexibility \cite{julie12}. Many IDPs do ``fold'' or become otherwise 
ordered upon binding to their folded partners~\cite{Dyson2005}. Recently,
an IDP was found to fold by phosphorylation as a regulatory 
switch~\cite{Bah2015}. But 
IDPs can also form ``fuzzy complexes'' \cite{fuzzy08}, i.e., they remain 
largely disordered even when they are functionally bound, as exemplified 
by the interaction between the IDP cyclin dependent kinase inhibitor Sic1 
and the ubiquitin ligase SCF$^{\rm Cdc4}$ \cite{nash01,borg07,tanja2010}.
Although IDP conformations are largely disordered, their ensemble
distributions are not random. Like globular proteins, their behaviors are 
governed by their amino acid sequences~\cite{PappuCOSB}. 
In this respect, functional IDPs 
entail a type of biomolecular self-organization that is 
perhaps even more challenging to comprehend physico-chemically than 
the self-assembly of globular proteins.

More recently, a previously unexpected function of IDPs was discovered. Some 
IDPs have been found to be the main constituents of functional membraneless 
organelles in the living cell, the assembly and disassembly of which is 
apparently
underpinned by liquid-liquid IDP phase separation~\cite{Brangwynne2009,Dundr10,Rosen12,McKnight12,Lee13,Hyman14,wright14,Seydoux2014,Nott15,ElbaumGarfinkle15,Alberti2015,Brangwynne15,Nott16,Rosen2016,RosenPappu2016}.
Such IDP phase behaviors are sequence-dependent~\cite{Nott15,RosenPappu2016}
and may be regulated posttranslationally such as by 
phosphorylation of germ plasm component proteins in P 
granules \cite{Seydoux2014} and by arginine methylation of Ddx4$^{\rm N1}$ 
organelles~\cite{Nott15}. Droplets of these IDP-rich phase are fundamentally 
different from amyloids~\cite{Dobson2014,Fuxreiter2016} and condensed phases
of folded globular 
proteins~\cite{Vlachy93,Sear1999,Prausnitz2004,Schroer11,Moeller14,Prausnitz2015,Kastelic15}.
Membraneless organelles include nuceloli, Cajal bodies, and stress granules.
They are a form of cell compartmentalization, creating their 
peculiar microenvironments~\cite{Nott16,Rosen2016} that play critical 
roles in cellular integrity, homeostasis, gene regulation and the cell cycle. 
Examples include the RNA and protein-rich P granules in the germ cell of
{\it Caenorhabditis elegans}~\cite{Brangwynne2009,Seydoux2014} and
the RNA-protein (RNP) stress granules induced by exposing human (HeLa) 
cells to arsenate~\cite{Alberti2015}. Because of the central role
of membraneless organelles in cell cycle regulation and thus the disease 
process of cancer, and that RNP aggregates are often associated with
neurodegenerative diseases, a better understanding of the biophysics of 
IDP phase separation is not only of fundamental biological interest but 
also of tremendous medical relevance.

Unlike globular proteins, the role of hydrophobic interactions~\cite{Dill90}
are significantly less dominant in IDPs, which generally contains
fewer nonpolar but more polar, charged, and aromatic residues than
globular proteins (\cite{Simon2005,Uversky2010,Uversky2015} and references
therein). Many IDPs may be regarded as polyampholytes --- heteropolymers 
with both positively and negatively charged monomer 
units~\cite{Higgs91, Dobrynin04}. Dimensions of individual IDP molecules 
are affected not only by their total charges~\cite{Muller10} but also
sequence arrangement of charges~\cite{Das13}; their response to denaturant is
different from that of unfolded globular proteins \cite{jianhui15}.
Liquid-liquid phase separation of IDPs is sensitive to the charge pattern 
along their chain sequences as well~\cite{Quiroz15}. This is evident from 
recent experiments on the DEAD-box RNA helicase Ddx4~\cite{Nott15} and the 
Nephrin intracellular domain \cite{RosenPappu2016} showing different phase 
behaviors from their respective charge-scrambled 
mutants~\cite{Nott15,RosenPappu2016}.
These behaviors may in principle be addressed by explicit-chain 
simulations~\cite{Ruff15}. For computational tractability 
as well as conceptual advances, however, it is useful to develop 
an analytical theory of polymer solutions for IDP phase behaviors, 
notwithstanding the approximate and simplistic nature of all 
analytical theories of complex systems~\cite{Prausnitz2015}.
In view of the experimental observations, a desirable
feature of such theories is to afford an account of 
sequence-dependent long-range electrostatic interactions that goes 
beyond classical Flory-Huggins (FH) theory~\cite{FH1,FH2,Flory53}, as 
the latter is strictly speaking only suited for short-range, contact-like 
interactions such as those arising from hydrophobicity \cite{chandill91}.

A variety of analytical theories for electrolytes and charged polymers 
have been put forth during the last nearly one century since the 1923 
publication of the Debye-H\"{u}ckel (DH) theory that accounts for
the nonideality of electrolyte solutions~\cite{DH1923}.
DH theory, or equivalently the linearized Poisson-Boltzmann (PB) equation, 
has been applied to electrolyte and polyelectrolyte systems with 
identical or nonidentical ions~\cite{DHmodel, Vlachy99}. 
The PB equation has since been augmented, e.g., by incorporating
van der Waals forces, to construct more enriched theories 
such as that of Derjaguin, Landau, Verwey, and Overbeek (DLVO theory, 
1941, 1948) to rationalize the stability of lyophobic (solvent averting) 
colloid~\cite{Derjaguin41-DLVO, VO1948}. 
Subsequently, theories that go beyond the mean-field PB equation
such as the closure relations to solve the Wertheim-Ornstein-Zernike
(WOZ) equation~\cite{Friedman85, Wertheim86,Vlachy99}, e.g. the
hypernetted-chain/mean-spherical approximation (HNC/MSA)
(\cite{Blum77} and references therein)
and the Percus-Yevick approximation~\cite{Wertheim63}, as well as the
random phase approximation (RPA)~\cite{deGennes79, Borue88}, were
devised to model various electrolyte systems, including globular 
polyions as well as flexible polyelectrolytes and 
polyampholytes~\cite{Wertheim86, Rescic90, Vlachy99, Borue90, 
Gonzalez94, Mahdi00, Bernard00, Jiang01}.

Although the main concern of many of these theories has been
uniformly charged polyelectrolytes, the charge pattern on flexible 
polyampholytes has not escaped recognition as having a significant
impact on the conformational distribution of individual chains as well
as multiple-chain phase properties~\cite{Higgs91, Dobrynin04}. 
Based on beyond-DH formulations that take into account chain connectivity,
theories such as RPA~\cite{Wittmer93, Gonzalez94, Dobrynin95} and 
HNC/MSA~\cite{Bernard00, Jiang01, Jiang06} have been applied to 
study polyampholyte systems. However, such effort has not been
widely pursued, likely because of a lack of experimental impetus. 
In the absence of a chemical process comparable to the cellular apparatus 
capable of synthesizing specific amino acid sequences accurately,
synthesis of nonbiological polyampholytes with specific sequences
is extremely difficult if not impossible. In such synthesis, often
only the initial conditions can be controlled, resulting in a polymerization 
process that can only be monitored at the level of thermal 
average~\cite{Dobrynin04}. Consequently, research on 
nonbiological polyampholytes has focused on either the ensemble average of 
all possible random sequences~\cite{Higgs91, Wittmer93, Dobrynin95} 
or simple block polyampholytes with a strictly alternating~\cite{Wittmer93}, 
diblock~\cite{Gonzalez94}, or four-block charge 
patterns~\cite{Cheong05, Jiang06}. 

As far as charge effects in biomolecules are concerned, the Overbeek-Voorn 
(OV, 1957) theory~\cite{Overbeek57} is a rudimentary approach 
that combines DH theory with FH conformational entropy~\cite{Flory53, 
deGennes79}. 
OV theory has been applied to rationalize behavior of various complex
coacervations, e.g. between albumin and acacia~\cite{Schmitt98} as
well as between whey proteins and gum arabic~\cite{Weinbreck03,Schmitt11}.
Beyond-mean-field
theories have also been applied to model phase separation of {\it folded} 
globular proteins in aqueous solutions, wherein protein molecules are 
treated as spheres with no or few internal degrees of freedom, similar to the 
idealized folded states in earlier mean-field models for electrostatic 
effects in protein folding and stability~\cite{Stigter1991,Stigter1995}.
These include, but are not limited to, RPA~\cite{Vlachy93}, perturbation 
theories~\cite{Sear1999,Prausnitz2004} and, notably, 
a recent application Wertheim's thermodynamic
perturbation theory (TPT1)~\cite{Wertheim84}
by Vlachy and coworkers to rationalize protein aggregation 
induced by salt as well as the liquid-liquid co-existence curve of folded
lysozyme and $\gamma$ IIIa-crystallin solutions~\cite{Kastelic15}. One
of the goals of these studies has been to better understand crystallization 
of globular proteins for X-ray crystallography, including under harsh 
conditions with non-physiological pH and salt concentrations~\cite{Zhang09}. 
Effects of specific amino acid sequences were not considered in these
approaches.

The recent discovery of biologically functional IDPs as individual
molecules and also collectively via phase-separation has
sparked a renewed interest in polymer solution 
theories \cite{Hyman14,Brangwynne15,Ghosh15,Vovk16}.
For instance, Sawle and Ghosh developed an analytical
formulation to account for sequence-specific electrostatic 
effects on the dimension of individual polyampholytes \cite{Ghosh15},
providing predictions consistent with prior atomic simulation 
results \cite{Das13}. In this context, we recently 
outlined an approach to apply RPA theory to model sequence-specific 
long-range electrostatic effects in IDP phase separation~\cite{Lin16}. 
Compared to the mean-field FH~\cite{Lee13, Hyman14, Nott15}, DH and 
OV~\cite{Brangwynne15} theories that have been applied or advocated for
the study of IDP phase separation, our approach has the advantage
of treating chain connectivity rather explicitly and hence it allows for 
a direct, unambiguous input of the charge pattern along the chain 
sequence into the theory~\cite{Borue88,Gonzalez94,Mahdi00}. The
approach has been applied to the 236-residue IDP fragment Ddx4$^{\rm N1}$ 
of Ddx4, a protein required for the assembly and maintenance of 
membraneless organelles that are essential for germ cell development
in mammals, worms and flies~\cite{Nott15, Liang94}. 
Our theory is successful in rationalizing the experimentally observed
salt-dependent phase separation of Ddx4$^{\rm N1}$ as well as the
drastically different phase behaviors of Ddx4$^{\rm N1}$ and a charge
scrambled mutant Ddx4$^{\rm N1}$CS~\cite{Lin16}. The new results 
reported below are further development of this theory, including 
detailed comparisons with mean-field FH and OV/DH approaches, and 
exploration of extensions of the theory that may provide
deeper physical insights into
the fascinating phenomenon of biologically functional IDP liquid-liquid
phase separation in general.


\section{A sequence-specific RPA theory for polypeptide 
charge patterns}~\label{sec:seq-specified_RPA}

As described recently~\cite{Lin16}, our theory is for aqueous solutions
of neutral or nearly-neutral polyampholytes such as Ddx4$^{\rm N1}$ with 
small monovalent counterions and salt, and is based on previous RPA 
methods~\cite{Wittmer93, Gonzalez94}. (For simplicity of notation,
Ddx4$^{\rm N1}$ and its mutant are sometimes referred to simply as ``Ddx4'' 
in the discussion below when the meaning is obvious from the context).
Each polyampholyte chain is composed of $N$ amino acid residues (monomers) with 
charges  $\{\sigma_i\} = \{ \sigma_1,\sigma_2,\dots,\sigma_N \}$ given
in units of the electronic charge $e$ ($\sigma_i = \pm 1$ or $0$). 
Using the same notation as 
in~\cite{Lin16}, $\rho_m$, $\rho_c$, and $\rho_s$ are, respectively,
the average number densities of the monomers of the polyampholytes, 
counterions, and salt in a total solution volume $V$, where 
$\rho_c = \rho_m\left|\sum_i \sigma_i\right|/N$ because the number of 
counterions is equal to the total net charge of polyampholytes.

A major part of the configurational entropy of the system is based 
on the FH lattice model~\cite{Flory53, deGennes79}, in which spatial 
volume is partitioned into lattice sites each with a volume $a^3$, which
is of order of an individual solvent molecule; thus the total number
of lattice sites is $M = V/a^3$. The total free energy $F$ per lattice site
in units of $k_{\rm B}T$ is given by
\begin{equation}
f \equiv \frac{F a^3}{V k_{\rm B}T} = -s + f_{\rm int},
\label{first_eq}
\end{equation}
where $k_{\rm B}$ is Boltzmann's constant and $T$ is absolute temperature, 
$-s$ is the entropic contribution to free energy from FH consideration,
and $f_{\rm int}$ accounts for the effective solvent-mediated
interactions in the system. The $f_{\rm int}$ term is modeled 
under an RPA framework \cite{Lin16}, which provides a beyond-mean-field,
approximate account of local density fluctuations. RPA may be applied to 
any form of two-body interactions in principle. In this paper, however, 
we are interested in situations of polyampholyte phase separation where 
electrostatics is the dominant enthalpic contribution. Although
$f_{\rm int}$ is often characterized as ``enthalpic'' for terminological 
simplicity \cite{Lin16}, it is useful to keep in mind that effective 
solvent-mediated interactions such as hydrophobicity can be temperature 
dependent \cite{Dill89} and therefore contain entropic contributions.
As will be discussed below, the present RPA form of $f_{\rm int}$ also
contains entropic contributions from chain connectivity.

\subsection{Entropy in a size-dependent mean-field lattice model}~\label{sec:FH_entropy}

Most formulations of FH ~\cite{Flory53}, including
recent applications to IDP phase separation~\cite{Nott15,Lin16}, 
assume for simplicity that a solvent molecule is of the same size
of a monomer of the polymer of interest, each occupying a single lattice
site. Numerical results presented in this paper were obtained using
the same assumption. However, it is useful for future development of
theory to consider here a generalization of the FH approach that is
capable of accounting for solvent, monomers, salt ions, and counterions 
of different sizes.
Based on a detailed consideration of the physical meaning of the entropy
term in the FH formulation~\cite{chandill94} (especially discussion
relating to Figs.~4 and 5 in this reference), a generalized FH configurational
entropy of a collection of $m$ different types of polymers labeled by 
$i$ $=1,2,\dots, m$, each with $N_i$ monomers of size $r_i a^3$ ---
where individual solvent molecules, counterion and salt ions are regarded 
as special cases of polymers with $N_i=1$ --- may be derived as follows.
(Note that the meaning of index $i$ here is different from
that in some other parts of the paper where it is used to label
the monomers along a polymer. The meaning of ``dummy'' summation indices
should always be clear from the context nonetheless.) 
Without loss of generality, a value of $a^3$ can always be
chosen such that all $r_i$'s are integers or as close to being so as
desired. In such a system, the total number of lattice sites is given by
\begin{equation}
M = \frac{V}{a^3} = \sum_{i=1}^m r_i n_i N_i.
	\label{eq:all_lattice_sites}
\end{equation}
where $n_i$ is the number of type-$i$ polymer molecules in the solution.
We may now define the number density of type-$i$ monomer in units 
of $1/a^3$ as
\begin{equation}
\phi_i = \rho_i a^3 = \frac{n_i N_i}{M}, 
	\label{eq:phi_i}
\end{equation}
which allows rewriting Eq.~(\ref{eq:all_lattice_sites}) as
\begin{equation}
\sum_{i=1}^m r_i \phi_i = 1.
	\label{eq:phi_normalize}
\end{equation}
Note that $\phi_i$ is not the volume fraction \cite{Lin16} when $r_i \ne 1$. 

Following the argument in the original derivation of FH~\cite{Flory53}, 
we separate configurational entropy $S = k_B \ln \Omega$ into a term
arising from the translational freedom of the first monomer (or center of 
mass) of each polymer ($\Omega_{\rm CM}$) and another term accounting for 
the conformational freedom of each polymer with a fixed position for the 
first monomer or center of mass ($\Omega_{\rm conf}$) such that 
$\Omega= \Omega_{\rm CM} \Omega_{\rm conf}$. The translational term
corresponds to the number of ways of arranging $\sum_i n_i$ monomers
each being the first monomer of a polymer, which equals
\begin{equation}
\Omega_{\rm CM} = \frac{M!}{(M-\sum_{i=1}^m n_i)!\prod_{i=1}^m n_i!}
	\label{eq:trans_entropy}
\end{equation}
when all $r_i=1$.
As for the conformational term, we first recall the standard derivation
for the case of $r_i = 1$ for all $i$ on a lattice with $z$ nearest
neighboring sites for each lattice site. The decreasing probability
of inserting an additional monomer without violating excluded volume,
as the lattice is filled, is taken into account in a mean-field manner: 
\begin{equation}
\begin{aligned}
\Omega_{\rm conf} & = (z-1)^{{\cal N}_1} 
\prod_{l=0}^{{\cal N}_2-1}\left(1- \frac{\sum_{i=1}^m n_i + l}{M}\right) \\
& = \frac{(z-1)^{{\cal N}_1}}{M^{{\cal N}_2}} \left( M-\sum_{i=1}^m n_i \right)!
\; ,
	\label{eq:conform_entropy_r_equal}
\end{aligned}
\end{equation}
where ${\cal N}_1 = {\cal N}_2 = \sum_i n_i(N_i-1)$ is the total 
number of lattice sites (monomers) to be inserted after the position of 
the first monomer of each polymer has been fixed by the procedure 
described by Eq.~(\ref{eq:trans_entropy}).

In the generalized system with different $r_i$'s, each mon\-omer in 
a type-$i$ polymer consists of $r_i$ lattice sites and is here assumed to
have no internal degrees of freedom \cite{chandill94}. The placement of 
each of these monomers on the lattice may be seen as a series of 
successive attempted occupation of $r_i$ neighboring lattice sites 
in a unique order (hence no $z-1$ factor) that is consistent with 
the shape of the monomer and without violation of excluded volume.

In order to account for the possible violation of excluded volume when
the second and other lattice sites representing the first monomer
of each polymer (when $r_i > 1$) are placed after the first site 
has been positioned via the process described in 
Eq.~(\ref{eq:trans_entropy}), the expression needs to be modified:
\begin{equation}
\begin{aligned}
\Omega_{\rm CM} & 
\rightarrow \left[\prod_{l=0}^{\sum_{i=1}^m n_i (r_i-1) -1}
\left( 1-\frac{\sum_{i=1}^m n_i + l}{M} \right)\right]\Omega_{\rm CM}  \\
& = 
\frac{(M-\sum_{i=1}^m n_i)!}{(M-\sum_{i=1}^m n_i r_i)! M^{\sum_{i=1}^m n_i(r_i - 1)}} 
\Omega_{\rm CM}  \; .
\end{aligned}
	\label{eq:trans_entropy_mod}
\end{equation}
The same consideration implies that the total number of lattice 
sites ${\cal N}_2$ to be filled in for the process
described by $\Omega_{\rm conf}$ in Eq.~(\ref{eq:conform_entropy_r_equal}) 
is now modified to 
\begin{equation}
{\cal N}_2 = \sum_{i=1}^m n_i ( r_i N_i -1) ,
\end{equation}
with a corresponding modification to Eq.~(\ref{eq:conform_entropy_r_equal}):
\begin{equation}
\Omega_{\rm conf} \rightarrow
\frac{(z-1)^{\sum_{i=1}^m n_i(N_i-1)}}{M^{\sum_{i=1}^m n_i r_i(N_i-1)}} 
(M-\sum_{i=1}^m n_i r_i)! .
	\label{eq:conform_entropy_r_diff}
\end{equation}
By combining Eqs.~(\ref{eq:trans_entropy}), (\ref{eq:trans_entropy_mod}), and 
(\ref{eq:conform_entropy_r_diff}), we arrive at
the expression for the total number of FH microstates for
the generalized case:
\begin{equation}
\Omega = \Omega_{\rm CM} \Omega_{\rm conf} 
= \frac{M!}{\prod_{i=1}^m n_k !}  
\frac{(z-1)^{\sum_{i=1}^m n_i(N_i-1)}}{M^{\sum_{i=1}^m n_i(r_i N_i-1)}}.
\end{equation}
Thus the negative configurational entropy per lattice site 
\begin{equation}
\begin{aligned}
-s = & -\frac{1}{M} \ln \Omega = \sum_{i=1}^m 
\frac{n_i}{M} \ln n_i + \text{(terms linear in $n_i$)} \\
= & \sum_{i=1}^m \frac{\phi_i}{N_i} \ln \phi_i + \text{(terms linear 
in $\phi_i$)}
\end{aligned}
\end{equation}
is obtained
by applying Stirling's approximation $\ln x!\approx x\ln x - x$
for $x \gg 1$ (in which case the percentage error of omitting
the $\ln\sqrt{2\pi x}$ term in the approximation is negligible).
It follows that the mixing entropy is given by
\begin{equation}
\begin{aligned}
\Delta s(\{ \phi_i \} ) = & s(\{ \phi_i \} ) \!-\! 
\sum_{i^\prime = 1} r_{i^\prime} \phi_{i^\prime} 
s(\{ \phi_{i^\prime} \!=\! 1/r_{i^\prime},\phi_{i\neq i^\prime} \!=\! 0\} ) \\
= & \sum_{i=1}^m \frac{\phi_i}{N_i }  \ln(\phi_i r_i).
\end{aligned}
	\label{eq:s_mixing}
\end{equation}
As shown in \ref{app:binodal_general}, terms of the form $(\phi_i/N_i)\ln r_i$
in Eq.~(\ref{eq:s_mixing}) that are linear in $\phi_k$ have no effect 
on phase separation. Therefore, for application to phase separation, 
we may drop these terms and choose the totally demixed configuration as the 
reference state, then use the mixing entropy expression
\begin{equation}
-s =  \frac{\phi_m}{N} \ln \phi_m + \phi_c \ln \phi_c + 2\phi_s \ln \phi_s 
+ \phi_w\ln \phi_w
	\label{eq:entropy}
\end{equation}
as the generalized FH entropy term in Eq.~(\ref{first_eq})
for a system of polyampholytes, counterions, and salt. Here
$r_m$, $r_c$, and $r_s$, are size factors, respectively, for monomers of
the polyampholytes, counterions, and salt ions, $r_w$ and $\phi_w$ are,
respectively, the size factor and number density of water, satisfying
\begin{equation}
r_w\phi_w = 1-r_m\phi_m-r_c\phi_c-2r_s \phi_s
\label{incompressibility_eq}
\end{equation}
by virtue of Eq.~(\ref{eq:phi_normalize}).

\subsection{RPA treatment of electrostatics}

Derivation of RPA electrostatic energy of 
polyelectrolyte solutions is well 
documented~\cite{Borue88, Mahdi00, Ermoshkin04}. We include here
for completeness the steps we took to apply RPA to our
polyampholyte system, providing some details for the theory we outlined
recently \cite{Lin16}. 

RPA theory neglects all but the trivial zeroth-order and two-body correlation 
of density fluctuation. 
Consider a system of different types of monomers (labeled as 1,2,3, ...) with 
densities $\rho^{(1)}({\bf r}), \rho^{(2)}({\bf r}), 
\rho^{(3)}({\bf r}) ...$ that are functions of spatial position ${\bf r}$.
When the correlation among the densities is accounted for only up to
the two-body level, the partition function $Z$ can be expressed
as a path integral of all possible density profiles in terms 
of their Fourier-transformed fluctuations, viz.,
\begin{equation}
Z = \int {\cal D}|\rho_k\rangle \exp\left( -\frac{1}{2V}\sum_{k\neq0}\langle \rho_{-k} | \hat{\cal A}_k | \rho _{k} \rangle \right),
\label{path1}
\end{equation}
where $k$ here is the wave number (for all three dimensions of the
reciprocal space of any finite volume $V$), and all variables are 
Fourier transformed except $V$ which is the total volume of the 
solution system, appeared
here as the normalizing factor for $\sum_k$. The matrix $\hat{\cal A}_k$ 
accounts for two-body interactions 
between arbitrary Fourier transformed
densities $\rho_k^{(i)}$ and $\rho_k^{(j)}$ of monomer types $i$ and $j$;
$\hat{\cal A}_k$ may correspond to any interaction/correlation
that depends only on the relative position ${\bf r}-{\bf r}^\prime$ of 
$\rho({\bf r})$ and $\rho({\bf r}^\prime)$ in real space. 
Fourier transformations of density fluctuations are represented by
the column vector
\begin{equation}
| \rho_k \rangle_i \equiv \rho_k^{(i)} 
\label{eq:vector}
\end{equation}
and $\langle \rho_{-k} | = (| \rho_k \rangle^*)^{\rm T}$ is the 
conjugate row vector.
The path integral in Eq.~(\ref{path1}) is over all monomer types,
and over all wave numbers 
except $k=0$, i.e.,
\begin{equation}
{\cal D}| \rho_k \rangle = \prod_{i=1}^m\prod_{k\neq 0} d\rho_{k}^{(i)} \; .
\end{equation}
The $k=0$ term is omitted because 
\begin{equation}
\rho_{k=0}^{(i)} = \int d^3 r \rho^{(i)}(r) = n_i
\end{equation}
is the total number of type-$i$ monomers, which is a constant
that does not contribute to density fluctuation~\cite{DoiEdwardsbook}. 

In the absence of interactions among the monomers, $\hat{\cal A}_k$
accounts only for geometric constraints such as chain connectivity, in
which case $\hat{\cal A}_k$ is the inverse of the {\em bare} two-body 
correlation matrix $\hat{G}_k$, viz., 
\begin{equation}
{\cal A}_k = \hat{G}_k^{-1} \; .
\label{eq:A_and_invG}
\end{equation}
This relationship is readily verified by recalling that the two-body 
correlation is defined by the density conjugate field $h_k^{(i)}$ via
the partition function
\begin{equation}
\begin{aligned}
Z(| h_k \rangle) \!=  & \int {\cal D}|\rho_k\rangle e^{-\sum_{k\neq0} \left(\frac{1}{2V} \langle \rho_{-k} | \hat{\cal A}_k | \rho _{k} \rangle+\langle h_{-k} | \rho_k \rangle \right)} \\
= & \exp\left( \frac{V}{2} \sum_{k\neq0}\langle h_{-k} | \hat{\cal A}^{-1}_k |
  h_k \rangle \right).
\end{aligned}
\end{equation}
The correlation function itself is defined as
\begin{equation}
\langle \rho_{k}^{(i)}\rho_{-k}^{(j)} \rangle 
= V\left(\hat{G}_k\right)_{ij}, \label{eq:rho_rho=G}
\end{equation}
where the $V$ factor appears because of translational invariance. 
Since the correlation function can be
determined as the second-order derivative of $\ln Z(| h_k \rangle)$,
\begin{equation}
\langle \rho_{k}^{(i)}\rho_{-k}^{(j)} \rangle =
\frac{\partial^2 \ln Z}{\partial h_{k}^{(i)}\partial h_{-k}^{(j)}} 
\Biggr \vert_{h_{k}^{(i)}=h_{-k}^{(j)}=0}
= V\left(\hat{\cal A}^{-1}_k\right)_{ij} \; ,
	\label{eq:Z_dhdh=A}
\end{equation}
Eq.~(\ref{eq:A_and_invG}) follows from Eqs.~(\ref{eq:rho_rho=G}) 
and (\ref{eq:Z_dhdh=A}).

In our system of polyampholytes and monomeric ions, the matrix
elements of $\hat{G}_k$ can be classified into four categories:
bare monomer-monomer (MM), ion-ion (II), and monomer-ion (MI) 
correlations with their respective matrices $\GMM$, $\GII$, $\GMI$. 
Accordingly, $\hat{G}_k$ can be arranged as
\begin{equation} 
\hat{G}_k = \left(
\begin{array}{cc}
\GMM(k) & \GMI(k)  \\
\GIM(k) & \GII(k)
\end{array}\right),
\end{equation}
where $\GIM = \GMI^{\rm T}$. 
In general, all four correlation submatrices depend on the 
configurational constraints in the system. For example, crosslinking
in polymer gels~\cite{Ermoshkin03a,Ermoshkin04} or ion condensation in 
charged polymer systems~\cite{Ermoshkin03b} can result in non-trivial 
correlations. Here we restrict to a simple RPA model of charged polymers 
and ions~\cite{Wittmer93, Borue88, Mahdi00} 
that treats all molecules as Gaussian chains or monomeric particles with
no excluded volume. (Excluded volume is accounted for separately 
in a mean-field manner by the FH configurational entropy term.) 
In this case, correlation
exists only among monomers belonging to the same polymer because
of chain connectivity. Consequently, all matrix elements of 
$\GMI$ and $\GIM$ are zero. In other words, there are no monomer-ion 
correlations in the model. The ion-ion correlation is 
the sum of the self-correlation of all ions, given by 
the $2\times 2$ diagonal matrix 
\begin{equation} 
\GII(k) = \hat{\rho}_I  =
\left(
\begin{array}{cc}
\rho_+ & 0 \\
0 & \rho_-
\end{array}
\right),
\end{equation}
where $\rho_\pm$ are the densities of positive and negative ions, 
$(\rho_+, \rho_-) = (\rho_s\!+\!\rho_c, \rho_s)$ or
$(\rho_s, \rho_s\!+\!\rho_c)$ when the net polyampholyte charge is, 
respectively, negative or positive \cite{Lin16}.
Finally, the monomer-monomer correlation matrix is calculated 
as the product of the structure factor of a Gaussian chain and
polymer density,
\begin{equation}
\GMM(k) = (\rho_m/N)\GM(k),
\end{equation}
where the matrix elements of $\GM(k)$ is given by
\begin{equation}   
\GM(k)_{ij} =\langle e^{i \vec{k} 
\cdot (\vec{x}_i - \vec{x}_j )} \rangle = e^{-(kb)^2|i-j|/6},
\end{equation}
with $b$ being the length of polymeric links of the Gaussian chains, 
and $\langle ... \rangle$ representing average over all possible 
polymer conformations. The matrix element $(\GM(k))_{ij}$ accounts
for the correlation between the $i$th and $j$th monomers in 
a Gaussian chain~\cite{DoiEdwardsbook, Grosberg94, deGennes79}.
Taken together, the above considerations lead to a block diagonal form for
the bare correlation matrix:
\begin{equation} 
\hat{G}_k = \left(
\begin{array}{cc}
(\rho_m/N)\GM(k) &  0 \\
0 & \rho_I
\end{array}\right) \; .
\end{equation}

\begin{figure*}
\fbox{
\includegraphics[width=\textwidth]{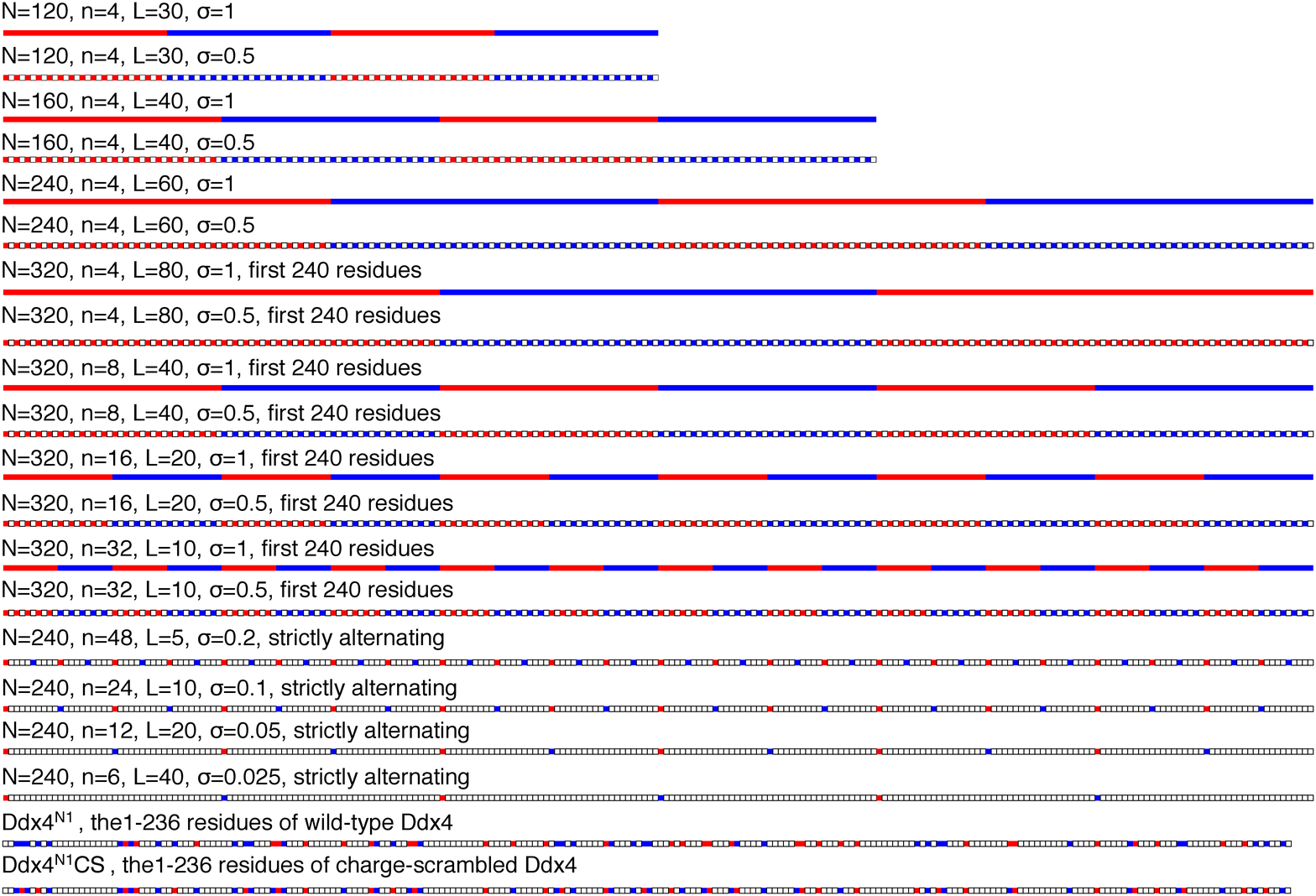}}
   \caption{Sequences studied in the present work. Monomers (residues) 
are depicted
as small squares. Each colored square is a monomer with 
a positive (red) or negative (blue) electronic charge. Neutral monomers 
are white. For the two Ddx4 sequences, red monomers 
correspond to either arginine (R) or lysine (K); blue monomers correspond
to either aspartic (D) or glutamic (E) acid \cite{Nott15, Liang94}. 
$N$ is chain length (number of monomers), $n$, the number of charged 
 blocks, $L$, length of an individual block, and $\sigma$,
the fraction of monomers that are charged.
The first 14 sequences are studied in Fig.~\ref{fig:ps_block_diff}, 
the next four sequences in Fig.~\ref{fig:strict_alter_sigdiff}, and 
the last two 
sequences in Figs.~\ref{fig:Ddx4_N1-N1CS_compare}, 
\ref{fig:RPA_and_RPA_FH}, and \ref{fig:three-model_compare}.
For the $N=320$ sequences, only the first 240 monomers are
included in this figure. The other 80 monomers of each of these 
sequences follow the same charge pattern of the first 240 
monomers shown.}
   \label{fig:seqs}
\end{figure*}

We now proceed to add inter-monomer interactions into the system. 
RPA assumes that the interactions are weak such that the Gaussian-chain 
geometry described in the bare correlation matrix $\hat{G}$ remains
a good approximation for the polyampholytes. In general, 
sequence dependence of conformational distribution, which is expected
physically \cite{Das13}, may be treated approximately by replacing the 
(bare) Kuhn length $b$ by a sequence-dependent renormalized Kuhn 
length \cite{Ghosh15}. For tractability, however, here we assume that 
$\hat{G}$ is sequence independent.  Under this assumption,
for any given interaction matrix $\hat{U}_k = \hat{u}_k/(k_{\rm B} T)$, 
its perturbative effect on the partition function 
can be calculated as the average 
of $\exp(-\hat{U}_k)$ over all Gaussian-chain configurations, 
\begin{equation}
\begin{aligned}
Z_u = & \frac{\int {\cal D}| \rho_k \rangle
\exp\left( -\frac{1}{2V} \sum_{k\neq0}\langle \rho_{-k} 
| (\hat{G}_k^{-1} + \hat{U}_k)  | \rho_k \rangle \right) 
}
{\int {\cal D}| \rho_k \rangle
\exp\left(  -\frac{1}{2V} \sum_{k\neq0}\langle \rho_{-k} 
| \hat{G}_k^{-1} |  \rho_k \rangle \right)
} \\
& \times\exp\left(  -\frac{1}{2V}  \langle \rho_{k=0} 
| \hat{U}_{k=0}  | \rho_{k=0} \rangle   \right)
	\label{eq:Zu_path_integral} \\
& \equiv Z_u^{(1)}Z_u^{(2)} \; .
\end{aligned}
\end{equation}

Since all eigenvalues of $\hat{G}^{-1}_k$ and $\hat{G}^{-1}_k+\hat{U}_k$ 
are positive for Coulomb interaction, the path integrals in 
Eq.~(\ref{eq:Zu_path_integral}) can be calculated by Gaussian 
integrations to yield
\begin{equation}
Z_u^{(1)} = \frac{\prod_{k\neq0} \sqrt{\det\left( \hat{G}_k^{-1} \!+\! \hat{U}_k \right)}}{\prod_{k\neq0} \sqrt{\det\left( \hat{G}_k^{-1} \right)}} = \prod_{k\neq0} \sqrt{\det\left(1 \!+\! \hat{G}_k \hat{U}_k \right) } \; .
\end{equation}
The $Z_u^{(2)}$ factor for $k=0$ depends on the interaction potential
$U$. For Coulomb interaction, the neutrality of the system 
requires~\cite{Ermoshkin04}
\begin{equation}
\sum_{ij} \rho_{k=0}^{(i)} U_{k=0}^{ij} \rho_{k=0}^{(j)} 
\propto  (\sum_i n_i q_i)^2 = 0,
\end{equation}
and thus
\begin{equation}
Z_u^{(2)} = \exp(0) = 1 \; .
\end{equation}
We hereafter consider $\fel$ for Coulomb interaction
as the $f_\text{int}$ term in Eq.~(\ref{first_eq}) 
and calculate it as the logarithm of $Z_u^{(1)}$ divided 
by the total number of lattice sites $V/a^3$:
\begin{equation}
\fel = -\frac{a^3}{V}\ln Z_u^{(1)} = 
\frac{a^3}{2V}\sum_{k\neq0} \ln[\det(1 + \hat{G}_k\hat{U}_k)] \; .
\label{fel_eq}
\end{equation}
In doing so, we are using $Z_u$ in Eq.~(\ref{eq:Zu_path_integral}) 
as the partition function of the system. 
This is appropriate for phase separation
studies because it amounts to choosing as reference state 
a noninteracting system that does not phase separate and whose 
partition function is given by the denominator of the first line 
of Eq.~(\ref{eq:Zu_path_integral}).
Approximating the summation in Eq.~(\ref{fel_eq}) as an integration, 
$(1/V)\sum_k \to \int d^3k/(2\pi)^3$~\cite{DoiEdwardsbook, Ermoshkin04}, 
and subtracting the self electrostatic energy of all charges to eliminate 
unphysical ``ultraviolet'' divergences at 
$k\!\to\!\infty$~\cite{Wittmer93, Ermoshkin03a, Ermoshkin04, QFT}, we arrive
at the RPA formula~\cite{Ermoshkin03a, Ermoshkin03b, Ermoshkin04}
\begin{equation}
f_{\rm el} = \frac{1}{2}\int \frac{d^3 (ka)^3}{(2\pi)^3} 
\left\{\ln[\det(1 + \hat{G}_k\hat{U}_k)] - \mathrm{Tr}
(\hat{\rho}\;\hat{U}_k) \right\}.
	\label{eq:fel_ori}
\end{equation}
As in \cite{Lin16}, here we consider a Coulomb potential with 
a short-range physical cutoff on the scale of monomer 
size~\cite{Ermoshkin03a, Ermoshkin04},
\begin{equation}
U(r) = \frac{l_B(1-e^{-r/b})}{r} \; ,
\end{equation}
where $r$ is spatial distance between electric charges.
In $k$-space, this becomes
\begin{equation}
\hat{U}_k = \frac{4\pi l_B}{k^2[1+(kb)^2]}| 
q\rangle \langle q | \equiv \lambda(k)| q\rangle \langle q |  \; ,
	\label{eq:Uk}
\end{equation}
where $l_B = e^2/(4\pi\epsilon_0\epsilon_r k_{\rm B} T)$ is the 
Bjerrum length, $\epsilon_0$ is vacuum permittivity, $\epsilon_r$ is 
relative permittivity (dielectric constant) governing the interaction, 
$| q \rangle$ is the column vector for the charges of the monomers 
and monovalent ions, and $\langle q | \equiv | q \rangle^{\rm T}$ is 
the transposed row vector, with components  $q_i = \sigma_i$ 
for $1 \leq i \leq N$, $q_{N+1} =1$, and $q_{N+2} = -1$. 
The determinant in Eq.~(\ref{eq:fel_ori}) can now be simplified as
\begin{equation} 
\begin{aligned}
\det(1 + \hat{G}_k\hat{U}_k) = & 1+ \lambda(k) 
\langle q | \hat{G}_k | q \rangle  \\
= & 1 + \lambda(k) \left( 2\rho_s \!+\! \rho_c \!+ 
\frac{\rho_m}{N}\! \langle \sigma | \GM(k) | \sigma \rangle  \right)
\end{aligned}
	\label{eq:det(1+GU)}
\end{equation}
by using Sylvester's identity $\det(\hat{I}_\mu + \hat{A}\hat{B})
=\det(\hat{I}_\nu + \hat{B}\hat{A})$ where $\hat{A}$ and $\hat{B}$
are, respectively, any $\mu\times\nu$ and $\nu\times\mu$ matrices
and $\hat{I}_\mu$ and $\hat{I}_\nu$ are, respectively, $\mu\times\mu$
and $\nu\times\nu$ identity matrices~\cite{Borue88}. 
The second term in Eq.~(\ref{eq:det(1+GU)}) is a linear function 
of $\rho_m$, and 
\begin{equation}
\ln[\det(1 + \hat{G}_k\hat{U}_k)] 
\xrightarrow{k\to\infty} \lambda(k) \langle q | \hat{G}_k | q \rangle.
\end{equation}
This limiting property allows us to replace
the trace in Eq.~(\ref{eq:fel_ori}) by 
$\lambda(k) \langle q | \hat{G}_k | q \rangle$ to facilitate
numerical integration for $\fel$~\cite{Mahdi00}. As will be discussed 
in Sec.~\ref{sec:mean-field_phase} and \ref{app:binodal_general}, 
the additional term linear in $\rho_m$ 
has no effect on phase separation.

As in \cite{Lin16}, a set of dimensionless variables are introduced
for our analysis. First, a reduced temperature is defined by rescaling 
temperature with electrostatic energy,
\begin{equation}
T^* \equiv {b}/{l_B} = {4\pi\epsilon_0\epsilon_r k_{\rm B} Tb}/{e^2} \; .
	\label{eq:Teq}
\end{equation}
In addition, the $\fel$ integration is rewritten in terms of the 
reduced wave number $\kr = kb$. We notice that $\GM(k)$ is 
a function of $kb$ and thus we rewrite it as $\GM(\kr)$. Together
with the rescaled densities $\phi_m = \rho_m a^3$, $\phi_c = \rho_c a^3$, 
and $\phi_s = \rho_s a^3$, $\fel$ may be expressed in
terms of dimensionless variables only, 
\begin{equation}
\fel = \int \frac{d\kr\kr^2}{4\pi^2} \left\{\frac{1}{\eta}\ln
\left[ 1 + \eta {\cal G}(\kr)\right] - {\cal G}(\kr) \right\} \; ,
	\label{eq:fel_dimless}
\end{equation}
where $\eta = (b/a)^3$ is the ratio between the cube of 
polymer link length and water molecular size when $r_w=1$, and
\begin{equation}
\!\!{\cal G}(\kr) = \frac{4\pi}{\kr^2[1+\kr^2] T^*}
	\left( 2\phi_s \!+\! \phi_c \!+\! \frac{\phi_m}{N}  
\langle \sigma | \GM(\kr) | \sigma \rangle \right). 
	\label{eq:G_cal}
\end{equation}
Using numerical integration of Eq.~(\ref{eq:fel_dimless}) for
$f_{\rm int}=\fel$ in Eq.~(\ref{first_eq}) and 
the FH configurational entropy expression in
Eq.~(\ref{eq:entropy}) for $-s$,
this formulation is applied below to study the phase behaviors of the
sequences in Fig.~\ref{fig:seqs}.

\begin{figure}[t]
\includegraphics[width=\columnwidth]{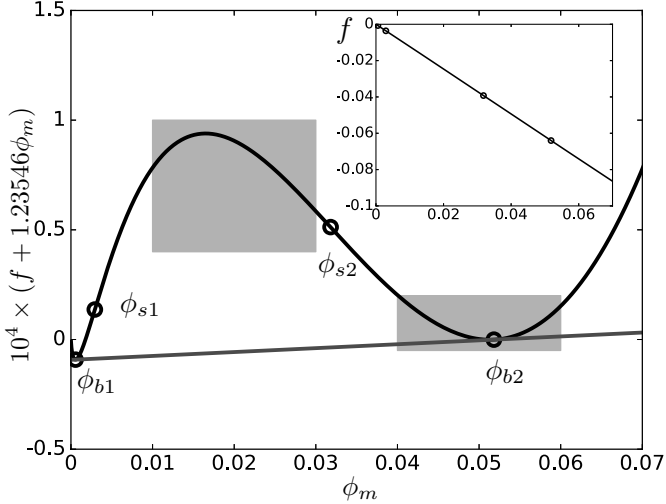} 
   \caption{The $f(\phi_m)$ function [Eq.~(\ref{first_eq})] 
at $T^*=1/3.5=0.286$ in our
RPA model for Ddx4$^{\rm N1}$ \cite{Lin16} {\it minus} $-1.23546\phi_m$,
where $-1.23546$ is the slope of the secant from $\phi_m=0$ to $0.05$. 
$f(\phi_m)$ itself is shown in the inset.
$\phi_{s1}$, $\phi_{s2}$ are the two spinodal points satisfying 
$f''_{el}(\phi_m) = 0$, whereas $\phi_{b1}$, $\phi_{b2}$ are the two binodal 
points sharing a common tangent. The $f$ values for these four 
$\phi_m$'s are marked by circles. The subtraction of a term linear
in $\phi_m$ from $f(\phi_m)$ serves to better exhibit its curvature,
which is less apparent visually without the 
subtraction (inset). The two shaded boxes
mark two regions of $f(\phi_m)$ that are further
analyzed, respectively, in Fig.\ref{fig:concave_convex}b and
\ref{fig:concave_convex}c.}
   \label{fig:f_curvature}
\end{figure}

\subsection{Determination of phase boundaries}~\label{sec:mean-field_phase}
\nobreak

\begin{figure*}[t]
\includegraphics[width=\textwidth]{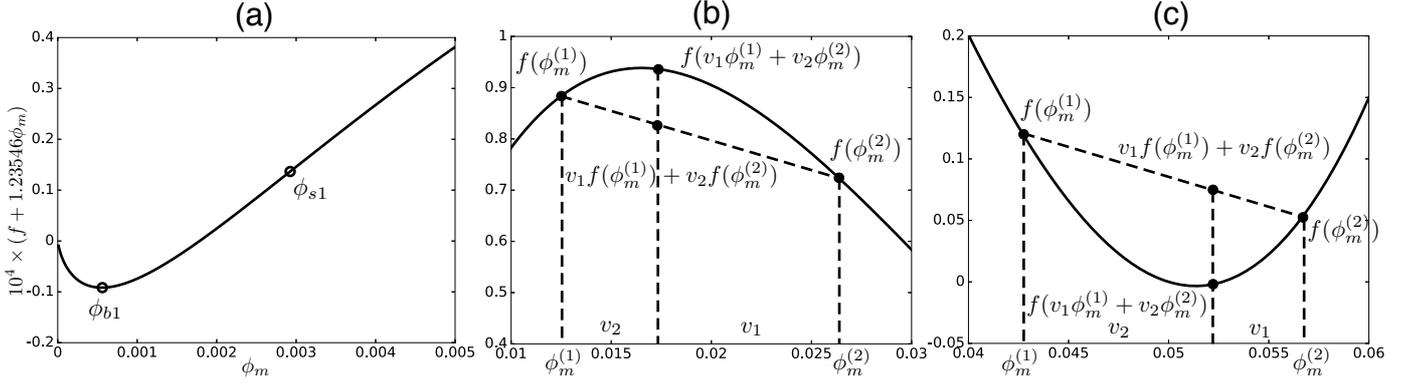}
   \caption{Phase stability of the RPA model for Ddx4 \cite{Lin16}. Shown
here are enlarged parts of the $f(\phi_m)$ function
in Fig.~\ref{fig:f_curvature}.
(a) Zoom-in display of the low-$\phi_m$ minimum of $f(\phi_m)$ near the 
origin. (b) The convex region of $f(\phi_m)$ satisfying 
Eq.~(\ref{eq:concave_cond}).
(c) The concave region near the second minimum of $f(\phi_m)$. This
region satisfies the inequality reverse to that 
of Eq.~(\ref{eq:concave_cond}).}
   \label{fig:concave_convex}
\end{figure*}

We first use the RPA model for Ddx4$^{\rm N1}$ with
$r_m=r_c=r_s=r_w=1$ \cite{Lin16} as an example 
to illustrate the procedure used to determine phase boundaries
(Figs.~\ref{fig:f_curvature} and \ref{fig:concave_convex}) 
and to highlight qualitative differences between RPA and FH
phase behaviors (cf. Fig.~4 of \cite{Hyman14}). The charge pattern
of Ddx4$^{\rm N1}$ is provided in Fig.~\ref{fig:seqs}.

Using the configurational entropy and interaction terms in 
Eqs.~(\ref{eq:entropy}) and (\ref{eq:fel_dimless}), the free energy
of Ddx4$^{\rm N1}$ solution is a function of polyampholyte and salt densities,
\begin{equation}
f = f( \phi_m, \phi_s) \; .
	\label{eq:f_phim+phis}
\end{equation}
Whether concentrations $( \phi_m, \phi_s)$ result in phase separation 
depends on the local curvature of $f$ on the $\phi_m-\phi_s$ plane. 
We consider only the salt-free ($\phi_s = 0$) case in 
Figs.~\ref{fig:f_curvature} and \ref{fig:concave_convex}.
The function $f$ exhibits two local minima (Fig.~~\ref{fig:f_curvature}), 
one at $\phi_m\approx 0.005$ (Fig.~\ref{fig:concave_convex}a), the
other at $\phi_m \approx 0.052$ (Fig.~\ref{fig:concave_convex}c). 
Two points in Fig.~\ref{fig:f_curvature}, 
$\phi_m=\phi_{s1}$, $\phi_{s2}$, satisfy 
\begin{equation} 
f'' \equiv \frac{d^2 f}{d \phi_m^2} = 0 \; .
	\label{eq:ddf=0}
\end{equation}
The free energy is a concave function ($f'' < 0$) between these two points.
Consider two concentrations $\phi_m^{(1)} < \phi_m^{(2)}$ within this
region. Any $\phi_m$ in between these two concentrations, i.e.,
$\phi_m^{(1)}\le \phi_m\le \phi_m^{(2)}$, can be written as  
\begin{equation}
\phi_m = v_1 \phi_m^{(1)} + v_2 \phi_m^{(2)} \; ,
\end{equation}
where the weight factors $v_1$,$v_2$ satisfy $0\le v_1, v_2 \le 1$.
A homogeneous concentration $\phi_m$ is not stable
in the concave region because its free energy is higher than
the free energy $\bar{f}$ of two
phase-separated (demixed) concentrations $\phi_m^{(1)}$ and $\phi_m^{(2)}$,
\begin{equation}
\bar{f} = v_1 f(\phi_m^{(1)}) + v_2 f(\phi_m^{(2)}) 
<  f(v_1 \phi_m^{(1)} + v_2 \phi_m^{(2)}) \; ,
	\label{eq:concave_cond}
\end{equation}
as can be seen in Fig.~\ref{fig:concave_convex}b by comparing the
free energy values along the secant from $\phi_m^{(1)}$ to 
$\phi_m^{(2)}$ and $f$ itself. In contrast, in a region with
$f'' > 0$, the homogeneous phase is seen to be 
more favorable by using the same consideration 
(Fig.~\ref{fig:concave_convex}c). The $f'' < 0$ condition defines the 
{\em spinodal} phase separation region in which the demixed state
is globally more favorable. For systems with more than two 
relevant molecular components, i.e. $f = f(\{ \phi_k \})$, the 
spinodal condition Eq.~(\ref{eq:ddf=0}) is generalized to
\begin{equation}
\det{\hat{\cal F}} = 0 \; , \; \hat{\cal F}_{ij} = 
\frac{\partial^2 f(\{ \phi_k \})}{\partial \phi_i \partial \phi_j} \; .
\end{equation}

As is conventional, the phase boundaries in \cite{Lin16} are
coexistence curves satisfying the {\em binodal} condition, which
corresponds to the equality of chemical potentials across phase
boundaries, and therefore may be determined graphically by constructing 
a common tangent (Fig.~\ref{fig:f_curvature}) -- because chemical 
potentials are by definition derivatives of free energy with respect 
to number of molecules. Specifically, the binodal concentrations 
$(\phi_m^\alpha, \phi_m^\beta)$ for two separate phases $\alpha,\beta$
is determined by the conditions
\begin{subequations}
\begin{align}
f'(\phi_m^\alpha) = & f'(\phi_m^\beta) \; , \\ 
f(\phi_m^\alpha) - \phi_m^\alpha f'(\phi_m^\alpha) = & f(\phi_m^\beta) - 
\phi_m^\beta f'(\phi_m^\beta),
\end{align}
	\label{eq:phm_separate}%
\end{subequations}
where $f'(\phi_m) = \partial f/\partial \phi_m$. 
These two equations 
uniquely define the common tangent of $f(\phi_m)$ at $\phi_m^\alpha$ 
and $\phi_m^\beta$~\cite{Voorn56b, Hyman14, Rubinstein03}, as 
illustrated by the example in Fig.~\ref{fig:f_curvature} with 
$\phi_m^\alpha$ and $\phi_m^\beta$ corresponding to
$\phi_{b1}$ and $\phi_{b2}$. 
While the spinodal region necessitates global phase separation
across the entire solution, phase separation can occur locally
in binodal regions (e.g., for bulk concentration $\phi_m$ in the 
regions $\phi_{b1} < \phi_m < \phi_{s1}$ and $\phi_{s2} < \phi_m < \phi_{b2}$
in Fig.~\ref{fig:f_curvature}). In these regions, strong local fluctuations 
into the spinodal region can be metastable, resulting in small liquid droplets 
with a different concentration. Generalizations of the bionodal 
conditions in Eq.~(\ref{eq:phm_separate}) are provided 
in \ref{app:binodal_general}.

It should be noted that salt concentration $\phi_s$ is taken to be 
uniform (constant across phase boundaries) here and in \cite{Lin16}. 
This need not be the case in general. Indeed, two-phase electrostatic 
coacervation in systems consisting of two, three, and four types of 
charged molecules have been well studied under the framework 
of DH theory~\cite{Voorn56a,Voorn56b,Voorn56c,Voorn56d}. 
Previous research, however, suggests that change in salt concentration 
is insignificant upon biological coacervation~\cite{Voorn56c, Overbeek57}.
We therefore assume for simplicity that the phase behavior of our model 
polyampholyte system does not involve fluctuation in salt concentration.

\section{Phase behaviors of salt-free block polyampholytes}~\label{sec:block}

To better understand the physical consequences of our RPA model and
to delineate its differences with FH theory in detail,
we extend our previous analysis \cite{Lin16} of simple salt-free ($\rho_s = 0$)
solutions of polyampholytes consisting of $n$ alternating charge blocks
(labeled by $\alpha$, $\beta=1$, 2, $\dots, n$) with length $L \!=\! N/n$.
The charge density in each block is $\sigma_\alpha^\text{block}
= (-1)^{\alpha-1}\sigma$, i.e., a charge per $1/\sigma$ monomers
(Fig.~\ref{fig:seqs}).
For simplicity, we assume, as before \cite{Lin16}, that all monomers
are of identical size, i.e.  $r_m = r_c = r_s = 1$, and the polymer 
link length is equal to the solvent length scale, i.e., $b =a$ 
[$\eta = 1$, Eq.~(\ref{eq:fel_dimless})].

The correlation matrix $\GM(\kr)$ for these highly regular charge patterns
is reduced to an $n\times n$ matrix for the blocks, $\GM^\text{block}(k)$. 
The diagonal terms of $\GM^\text{block}$, 
\begin{equation}
\GM^\text{block}(k)_{\alpha\alpha}
= \sum_{i,j=1}^L \sigma_i \sigma_j e^{-(kb)^2|i-j|/6} \; ,
\end{equation}
account for intra-block correlations.
Because only the $(1+l/\sigma)$th ($l = 0, 1, 2, ...$) residues are charged, 
as illustrated by the second sequence in Fig.~\ref{fig:seqs},
the summation of $\GM^\text{block}$ becomes
\begin{equation}
\GM^\text{block}(k)_{\alpha\alpha}
= \sum_{i,j=1}^{L\sigma} e^{-(kb)^2|i-j|/(6\sigma)},
\end{equation}
in which the $i$ and $j$ labels only go through the charged residues. The
$\GM^\text{block}(k)_{\alpha\alpha}$ is then calculated in terms of
$\expkR \equiv$ $\exp(-(kb)^2/(6\sigma))$ as
\begin{equation}
\begin{aligned}
\GM^\text{block}(k)_{\alpha\alpha} = & L\sigma
+ 2\sum_{j=1}^{L\sigma} \sum_{i=j+1}^{L\sigma} \expkR^{i-j} \\
= & \frac{1+\expkR}{1-\expkR} L\sigma
- \frac{2\expkR(1-\expkR^{L\sigma}) }{(1-\expkR)^2} \; ,
\end{aligned}
\end{equation}
where the $L\sigma$ term corresponds to the self-correlation of each
charged monomer. The off-diagonal terms account for inter-block correlations,
of which the lower triangular components ($\alpha > \beta$) are
\begin{equation}
\begin{aligned}
& \GM^\text{block}(\kr)_{\alpha>\beta} =
\sum_{i=(\alpha-1)L\sigma+1}^{\alpha L\sigma}
\sum_{j=(\beta-1)L\sigma+1}^{\beta L\sigma} \expkR^{i-j}  \\
= & \frac{  \expkR^{(\alpha-1)L\sigma+1} (1-\expkR^{L\sigma} ) }
{1-\expkR}
\times \frac{ \expkR^{-(\beta-1)L\sigma-1}(1-\expkR^{-L\sigma} ) }
{1-\expkR^{-1} } \\
= & \frac{\expkR}{(1-\expkR)^2} \expkR^{(\alpha-\beta-1)L\sigma}
(1-\expkR^{L\sigma})^2
\end{aligned}
\end{equation}
and the upper triangular part is given by switching $\alpha$ and $\beta$,
resulting in the general expression
\begin{equation}
\GM^\text{block}(k)_{\alpha\beta} =  \frac{\expkR}{(1-\expkR)^2}
\expkR^{(|\alpha-\beta|-1)L\sigma}(1-\expkR^{L\sigma})^2 \; .
\end{equation}


In the same vein, $| \sigma \rangle $ is grouped into $n$ components with $\sigma_\alpha  = (-1)^{\alpha-1}$. The RPA contribution from $\GM$ is then calculated by summing all $\GM^\text{block}(k)$ elements with alternating $\pm 1$ weight factors, yielding
\begin{equation}
\begin{aligned}
& \langle \sigma | \GM(k) | \sigma \rangle \\
& = n\GM^\text{block}(k)_{\alpha\alpha} + 2\sum_{\beta=1}^n 
			\sum_{\alpha=\beta+1}^n 
			(-1)^{\alpha-\beta}  \GM^\text{block}(k)_{\alpha\beta} \\ 
& = \frac{1+\expkR}{1-\expkR} n L \sigma - \frac{2\expkR}{(1-\expkR)^2}(1-\expkR^{L\sigma} ) n \\
	& \quad + \frac{2\expkR^{1-L\sigma}}{(1-\expkR)^2}(1-\expkR^{L\sigma})^2 
				\sum_{\beta=1}^n \sum_{\alpha\!=\!\beta+1}^n 
					(-\expkR^{L\sigma})^{\alpha-\beta} \\
& = N\left[ \sigma\frac{1+\expkR}{1-\expkR}
	- \frac{1}{L}\frac{4\expkR}{(1-\expkR)^2} 
		\left( \frac{1-\expkR^{L\sigma}}{1+\expkR^{L\sigma}} \right) \right] \\
	& \quad + \frac{2\expkR}{(1-\expkR)^2} 
		\left(\frac{1-\expkR^{L\sigma}}{1+\expkR^{L\sigma}} \right)^2 (1-(-1)^n \expkR^{N\sigma}) \; .
\end{aligned}
	\label{eq:sGs}
\end{equation}

\begin{figure}[t]
\includegraphics[width=\columnwidth]{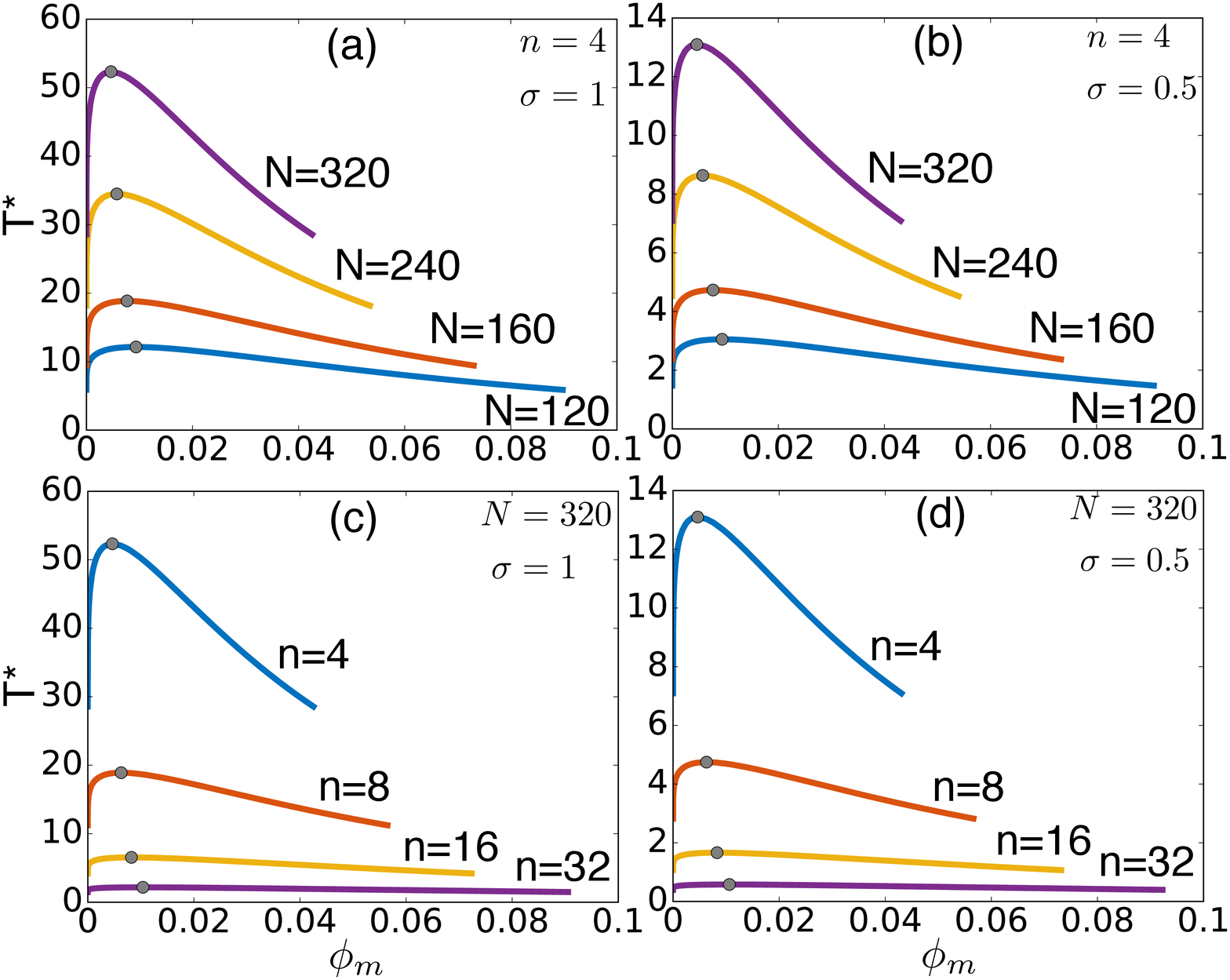} 
   \caption{Binodal phase boundaries for neutral salt-free polyampholytes 
with diffirent combinations of $N$, $n$, and $\sigma$: 
(a, b) $N=120$, 160, 240, and 320; $n = 4$. (c, d) $N = 320$, 
$n = 4$, 8, 16, and 32. (a, c) $\sigma = 1$.  (b, d) $\sigma = 0.5$.
The dots mark the critical points. Longer polyampholytes (larger $N$) with 
fewer blocks (smaller $n$) and higher charge densities (higher $\sigma$)
have stronger tendencies to phase separate.} 
   \label{fig:ps_block_diff}
\end{figure}

Here we begin by using these formulas to examine phase separation of 
neutral block polyampholytes ($n$ even and thus $\rho_c =0$). 
The general trend shown in Fig.~\ref{fig:ps_block_diff} is that
polyampholytes with greater $N$, fewer $n$, and greater $\sigma$ 
phase separate at higher temperature $T^*$.
Specifically, when the number of charge blocks, $n$, is fixed, the 
critical temperature $\Tscr$ (maximum $T^*$ 
of the phase boundary) increases whereas the critical volume fraction 
(concentration) $\phicr$ ($=\phi_m$ at $\Tscr$) decreases as the 
block length $L$ increases (Fig.~\ref{fig:ps_block_diff}a). 
This $\Tscr\to\infty$, $\phicr\to0$ trend as $N=nL\to\infty$ with 
fixed $n$ is consistent with previous reports for short diblock 
polyampholytes~\cite{Jiang06, Cheong05}. In the limit of $N\to\infty$, 
the first term in Eq.~(\ref{eq:sGs}), which diverges at $k\to 0$, 
dominates the integration for $f_{\rm el}$. As a result, $f_{\rm el}$ is 
in the form of the Eq.~(5) in Ref.~\cite{Wittmer93} with $\lambda=1$, 
proportional to $\phi_m^{3/2}$ as in Debye-H\"{u}ckel theory 
but with an infinitely large prefactor. In this limit, the block 
number $n$ becomes irrelevant and the system behaves just like 
a solution of diblock polyampholytes with $N\to\infty$, 
in which electrostatic interaction is so strong that the RPA is broken. 
Diblock polyampholytes are expected to form structural 
aggregates~\cite{Wittmer93, Borue90}. We note that this $f_{\rm el}$ is 
much stronger than the Debye-H\"{u}ckel theory for electrostatic 
coacervation that expects an $f_{\rm el}$ with a finite 
prefactor~\cite{Overbeek57}.

If the chain length $N$ is fixed, when $L$ decreases (i.e. $n$ increases), 
the first term in Eq.~(\ref{eq:sGs}) is unchanged, but the second term 
increases, and the third term slightly decreases. In the limit of $N\gg 1$, 
only the first two $O(N)$ terms are relevant, which yield a decreasing 
contribution to $\fel$ when $L$ decreases. As shown in 
Fig.~\ref{fig:ps_block_diff}, a smaller $\Tscr$ thus results from 
more but shorter charge blocks, qualitatively consistent with 
experiment on the charge-scramble mutant Ddx4$^{\rm N1}$CS~\cite{Nott15}. 
The increase of $n$ will ultimately arrive at the strictly alternating 
polyampholyte, in which the polyampholyte contribution to $f_{\rm el}$ 
integration is given by substituting $L=1/\sigma$ in Eq.~(\ref{eq:sGs}) 
to give
\begin{equation}
\frac{1}{N}\langle \sigma | \GM(k) | \sigma \rangle 
=  \frac{1-\expkR}{1+\expkR}\sigma + 
\frac{1}{N}\frac{2\expkR(1-(-1)^n \expkR^{N\sigma})}{(1+\expkR)^2} \; .
	\label{eq:sGs_alternating}
\end{equation}
In the limit of $N\to\infty$, only the first term in 
Eq.~(\ref{eq:sGs_alternating}) remains. As this term converges 
to $0$ when $k\to0$, the $f_{\rm el}$ behavior in low $\phi_m$ can 
be evaluated by expanding Eq.~(\ref{eq:fel_dimless}) using $\ln(1+x) 
\approx 1 + x - x^2/2! + ...$ to the second order of $\phi_m$, yielding
\begin{equation}
\begin{aligned}
\fel^\infty \approx & -\frac{(6\sigma)^{\frac{3}{2}}}{2} \int \frac{dk' k'^2 }{4\pi^2} \left( \frac{4\pi \phi_m }{6k'^2(1+ 6\sigma k'^2)T^*} \frac{1-e^{-k'^2}}{1+e^{-k'^2}}  \right)^2 \\
\equiv & -\chiel^{\infty}\phi_m^2
\end{aligned}
	\label{eq:alter_N_is_inf}
\end{equation}
with $k'^2 = (kb)^2/6\sigma$. The electrostatic energy now 
becomes a quadratic term of polymer concentration. Thus, the 
$N\to\infty$ alternating polymers are interacting in terms of an 
effective FH interaction with the FH $\chi$ parameter
\begin{equation}
\chiel^{\infty} = 
\frac{(6\sigma)^{\frac{3}{2}} }{18T^{*2}} 
\int \frac{dk'}{k'^2(1+6\sigma k'^2)^2}\left(  
\frac{1-e^{-k'^2}}{1+e^{-k'^2}} \right)^2 \; .
	\label{eq:chi_el}
\end{equation}
For weakly charged polyampholytes, $\sigma \ll 1$. Thus, the 
$(1+6\sigma'k'^2)$ factor in Eq.~(\ref{eq:chi_el}) can be approximated 
to $1$ as if the short range cutoff is irrelevant, resulting in
\begin{equation}
\chiel^\infty \approx 0.687 \sigma^{3/2} T^{*-2},
	\label{eq:chi_infty}
\end{equation}
which is equivalent to Eq.~(3) in Ref.~\cite{Wittmer93}.

\begin{figure}[t]
  \includegraphics[width=\columnwidth]{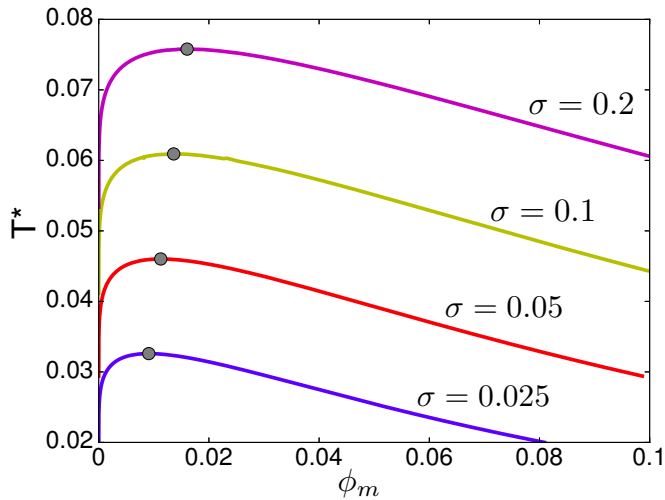} 
  \caption{Phase diagrams for strictly alternating polyampholytes 
with $N=240$ and $\sigma=0.025$, 0.05, 0.1, and 0.2, corresponding
to $n=6$, 12, 24, and 48, respectively. Critical points are marked
by the gray circles. Smaller $\sigma$ leads to lower $\phicr$ and lower $T^*$.}
	\label{fig:strict_alter_sigdiff}
\end{figure}

The critical point, however, does not obey the FH result that 
$\phicr \approx 1/\sqrt{N}$ and $\chiel^\infty = 0.5$. First, 
the second $O(1/N)$ term in Eq.~(\ref{eq:sGs_alternating}) enhances 
$\fel$, and thus $\phicr$ is smaller than the FH result. 
Even without considering this $1/N$ term in $\fel$, an approximate
description of these RPA results by a FH framework is still 
not applicable because FH does not take into account higher order terms 
of $\phi_m$. In the calculation of critical point, both the second 
and third order derivatives of free energy have to be zero. As FH 
only includes interactions up to $O(\phi_m^2)$, the third order 
derivative of interaction energy must be zero. The exact $\fel$ in 
RPA, however, can be expanded further,
\begin{equation}
\fel^\infty(\phi_m) = \fel^\infty(\phi_m\!=\!0) - \frac{1}{2}A \phi_m^2 + \frac{1}{6}B \phi_m^3 + O(\phi_m^4),
	\label{eq:fel_infty_phim3}
\end{equation}
where $A = 2\chiel^\infty$ and $B = \partial^3 \fel(\phi_m\!=\!0)/\partial \phi_m^3$  is given by
\begin{equation}
B = 2(6\sigma)^{\frac{3}{2}} \int \frac{dk' k'^2 }{4\pi^2} 
\left( \frac{4\pi }{6k'^2(1+ 6\sigma k'^2)T^*} 
\frac{1-e^{-k'^2}}{1+e^{-k'^2}}  \right)^3, 
\end{equation}
which in the limit of $\sigma \ll 1$ becomes
\begin{equation}
B \approx 1.187\sigma^{3/2} T^{*-3}.
	\label{eq:B+for_fel_3rd_div}
 \end{equation}
Noted that $\chiel^\infty = 0.5$ in Eq.~(\ref{eq:chi_infty}) would 
result in $T^* \sim \sigma^{3/4}$ and thus $B \sim \sigma^{-3/4} \gtrsim 1$, 
contradicting with the requirement of $B \approx 0$ in FH theory. 

Because the third order derivative of $\fel$ as well as the 
finite-size effect from $N$ and $\sigma$ are not negligible in our
RPA model, the FH inference based on Eq.~(\ref{eq:chi_infty}) does not 
lead to the correct critical point. In \ref{app:alter_N_infty} 
we describe a semi-analytic solution for critical point in the limit 
of $N\to\infty$ for any $\sigma$. Phase diagrams of a series of strictly 
alternating polyampholytes with identical $N$ are shown
in Fig.~\ref{fig:strict_alter_sigdiff}. 
In contrast to the decreasing $\Tscr$ and increasing $\phicr$ with
increasing number of charge blocks $n$ when charge density $\sigma$ is fixed 
(Fig.~\ref{fig:ps_block_diff}c,d), when the charge per block is fixed
as in Fig.~\ref{fig:strict_alter_sigdiff}, {\it both} $\Tscr$ and $\phicr$
increase with increasing number of charge blocks $n$. In other words,
whereas the variations of $\Tscr$ and $\phicr$ in Fig.~\ref{fig:ps_block_diff}
impact phase separation in the same direction, the variations of 
$\Tscr$ and $\phicr$ for strictly alternating polyampholytes in 
Fig.~\ref{fig:strict_alter_sigdiff} are such that they have opposite 
effects on the tendency of the solution to phase separate.

\section{Comparison of RPA and other analytical models for Ddx4 phase 
separation}

\subsection{The RPA model}

\begin{figure}[t]
  \includegraphics[width=\columnwidth]{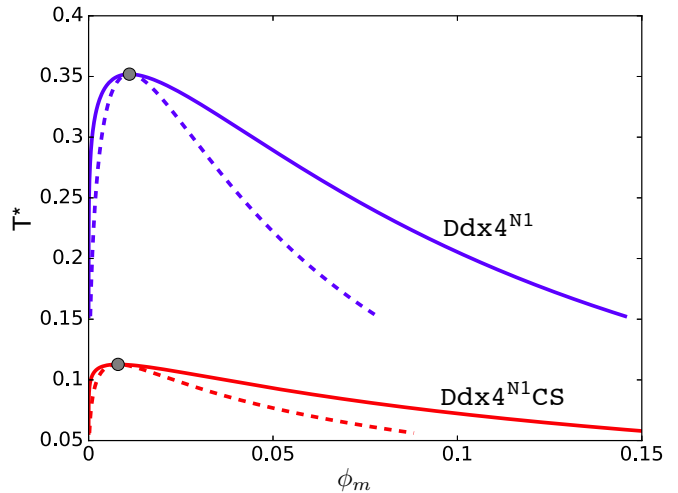}
  \caption{ 
 The salt-free phase diagrams for the two Ddx4 sequences in our RPA
theory. Solid lines are binodal boundaries that have been
shown in \cite{Lin16}, dashed lines are spinodal boundaries. 
 Gray circles mark the critical points. In this theory, $\Tscr$ of 
 Ddx4$^\text{N1}$CS is $\approx 1/3$ that of Ddx4$^\text{N1}$.}
	\label{fig:Ddx4_N1-N1CS_compare}
\end{figure}

We have applied our RPA theory to Ddx4 by considering the exact 
amino acid sequences of residues 1--236 of wildtype Ddx4 
(Ddx4$^\text{N1}$) and its charge-scrambled 
mutant (Ddx4$^\text{N1}$CS)~\cite{Nott15} 
(Fig.~\ref{fig:seqs})
to compute their binodal~\cite{Lin16} 
and spinodal boundaries (Fig.~\ref{fig:Ddx4_N1-N1CS_compare}). 
In this calculation, we assume for simplicity $r_m = r_c = r_s =  \eta = 1$. 
We have considered the general size-dependent formulation in 
Sec.~\ref{sec:FH_entropy} with reasonable variations in monomer sizes 
and found that the variations we tested only resulted in shifts of the 
phase boundaries in the vertical ($T^*$) direction without much 
alternation of their overall shapes.
Consistent with the experimental observation that 
Ddx4$^\text{N1}$ phase separates but
Ddx4$^\text{N1}$CS does not~\cite{Nott15}, 
Fig.~\ref{fig:Ddx4_N1-N1CS_compare} indicates that Ddx4$^\text{N1}$CS 
with its scrambled charged pattern has a critical temperature 
$T^*_{\rm cr}$ that is only about 1/3 that
of the wildtype in our RPA model \cite{Lin16}.
However, the critical concentration $\phicr$ of Ddx4$^\text{N1}$CS
is larger, not smaller, than that of the wildtype.
This trend of $(T^*_{\rm cr},\phicr)$ variation
is reminiscent of the $\sigma$ dependence of strictly alternating 
polyampholytes in Fig.~\ref{fig:strict_alter_sigdiff}.

\begin{figure*}[t]
  \includegraphics[width=\textwidth]{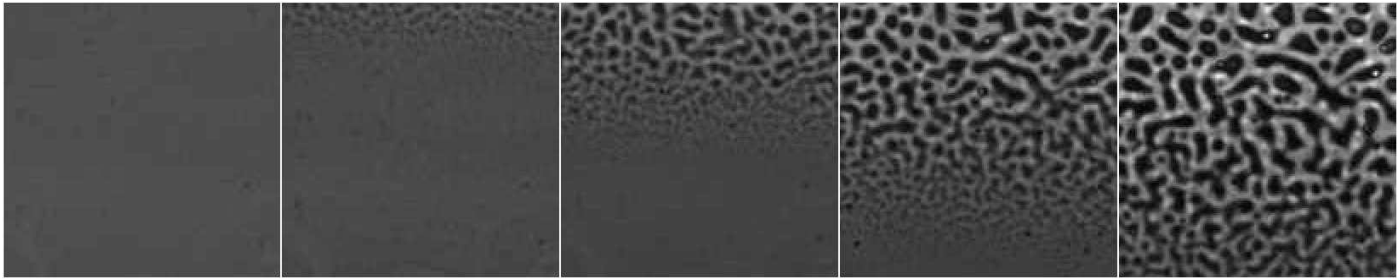} 
  \caption{Spinodal behavior of Ddx4. The montage shown was obtained
by drying a droplet of Ddx4 in aqueous buffer solution and recorded
using a 10x objective. The time lapse between successive frames (from 
left to right) was 0.32 second. The edge of the circular drying droplet 
was on the top of the frame though well out of view. Consequently, 
evaporation and hence increase in Ddx4 concentration was proceeding 
from top to bottom, resulting in the upper parts reaching the spinodal 
regime before the lower parts of the frame (courtesy Timothy J. Nott, 
unpublished).}
	\label{fig:Ddx4_spinodal}
\end{figure*}

The coexistence of two phases in Ddx4$^{\rm N1}$ solution has been
characterized experimentally by a dramatic change of turbidity
by lowering temperature -- presumably as
the system crosses the binodal boundary -- 
when numerous droptlets with relatively higher concentration of 
Ddx4 are formed (Fig.~3D in Ref.~\cite{Nott15}).
Global spinodal phase separation of Ddx4$^{\rm N1}$ solution
has also been observed by rapid increase in concentration,
probably quenching the system well into the spinodal regime. 
In this case, no spherical droplets 
were formed. Instead, the condensed regions extended and then merged 
into a giant network (Fig.~\ref{fig:Ddx4_spinodal}).

\subsection{Proposed analytical models for IDP phase separation}

We now proceed to compare models that have been either applied 
to \cite{Nott15,Lin16} or proposed for \cite{Brangwynne15}
the study of IDP phase separation. These include the FH~\cite{Nott15} 
and RPA~\cite{Lin16} models for Ddx4, and the OV/DH approach
advocated in a recent review on the polymer physics of intracellular 
phase transitions~\cite{Brangwynne15}.
All these models are mean-field at one level in that
they all incorporate the classical FH lattice derivation of 
configurational entropy described here in Sec.~\ref{sec:FH_entropy}, 
which considers only bulk polymer concentration and ignores all 
local density fluctuation. However, the way interactions
within and among polymers are treated is quite different 
in these models.

In both FH and OV/DH, the amino acid residues in polypeptides
are treated as independent monomers. Chain connectivity is
only accounted for in a mean-field manner in the derivation
of configuration entropy (the $(z-1)$ factors in 
Eq.~(\ref{eq:conform_entropy_r_equal}); but this factor is irrelevant 
to phase separation). But chain connectivity is neglected entirely 
in the interaction term, precluding these theories from addressing 
sequence-dependent interactions directly.
For example, the FH model for Ddx4 in \cite{Nott15} uses
a single overall FH $\chi$ parameter as an average interaction
strength. It does not distinguish between charged and uncharged 
residues. Moreover, the derivation of FH assumes that interactions
are of short-range, point-contact type~\cite{Flory53, deGennes79}. 
As such, the applicability of FH to long-range Coulomb interactions 
may be limited, as is evident from the difference in 
$\phi_m$ dependence between its interaction term and that of OV/DH
(see below). The OV/DH approach recognizes charged and neutral 
residues so as to apply electrostatic interaction only to the former, 
but it does not distinguish between positive and negative charges.
Only the total number of charged monomers, counterions, and 
salt ions can be parametrized \cite{DH1923}. Thus, in contrast
to the RPA approach that offers a direct treatment of charge
pattern along the polypeptide sequence \cite{Lin16}, effects of charge pattern 
are absent in FH and OV/DH models. The consequences of the assumptions
in these models are examined below by applying them to the salt-dependent 
phase behavior of Ddx4$^{\rm N1}$.

To compare the model results with experiment, we equate $a^3$ with 
the volume of a single water molecule in all calculation below. Accordingly,
the volume fraction of amino acid residues is
given by $\phi_m$ $=236$[Ddx4]/(55.5 M) because
[H$_2$O] = 55.5 M. We restrict our consideration to
salt concentrations $\phi_s=$ 0.0018, 0.0027, 0.0036, and 0.0054,
which correspond, respectively, to [NaCl]=100, 150, 200, and 
300 mM~\cite{Lin16} that have been investigated experimentally~\cite{Nott15}.
To faciliate comparison with the experimental phase diagrams in
Fig.~4 of \cite{Nott15}, special focus is placed on the range of 
[Ddx4]=0 -- 400$\mu$M.  

\subsection{Flory-Huggins (FH) model}
\label{FH_subsection}

\begin{figure*}[t]
\includegraphics[width=\textwidth]{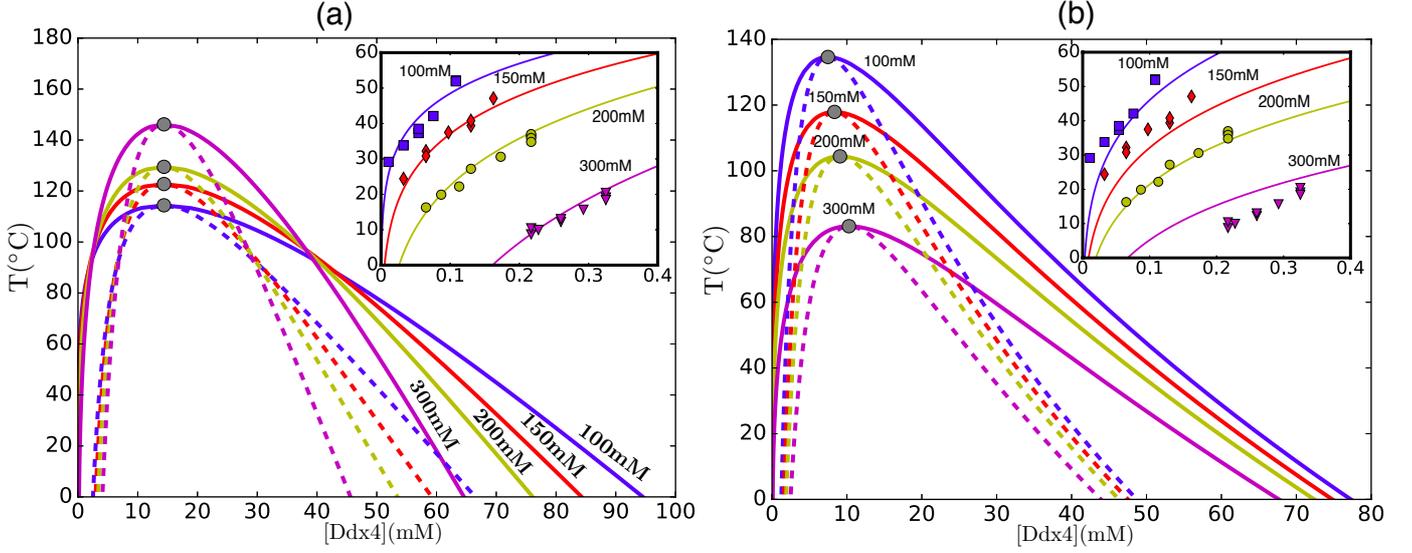}  
\caption{Salt-dependent phase diagrams of Ddx4$^\text{N1}$ 
predicted by different theories and their comparison 
with experimental [NaCl] dependence.
Binodal and spinodal boundaries are shown, respectively, as solid and
dashed lines. Critical points are marked by gray circles.
Experimental data~\cite{Nott15} are included in the insets for 
comparison (symbols; same color code).
(a) FH theory in Ref.~\cite{Nott15} as described in the Supplemental
Information of this reference. The fitting parameters are 
$\Delta H_0=78$ J mol$^{-1}$, $\epsilon_r = 50$, $r_d = 8.4354$\AA, 
$\Delta S_0 = -10.4922$ J mol$^{-1}$K$^{-1}$, and $B_0 = -0.19549$ (the
last two parameters are for Eq.~[S20] of \cite{Nott15} in the 
form of $\Delta S = \Delta S_0 + B_0 R \ln([{\rm NaCl}]/55.5)$,
where [NaCl] is in M). The critical point is always 
at $\phicr \approx 1/\sqrt{N}$ and 
$\chi_{\rm cr} \approx 0.5+1/\sqrt{N}$ because these values are
fixed in the FH model~\cite{deGennes79}. 
(b) DH theory given by $f_{\rm DH}$ in Eq.~(\ref{eq:fDH}); 
$\epsilon_r = 59.5$ is used to convert $T^*$ to $T$($^\circ$C) to fit
the experimental data.}
	\label{fig:Bd-FH_and_OV-DH}
\end{figure*}

FH model has been applied to analyze experimental
data of Ddx4 phase separation~\cite{Nott15}. 
This particular model supposes that Coulomb interaction contributes 
to attraction between all residue pairs, and the effect
of salt is a screening effect that decreases this general attraction. 
Following the description of the model in Supplemental
Information of \cite{Nott15}, we conducted the calculation anew,
arriving at a set of parameters that provide a reasonably
good fit to the experimental data (Fig.~\ref{fig:Bd-FH_and_OV-DH}a, inset).
Our fitted parameters were obtained by first fixing $\Delta H_0$ and 
$\epsilon_r$ to values consistent with those in \cite{Nott15}. The
resulting parameters differ from theirs, however, probably because 
there can be more than one optimum for such a multiple-parameter 
fit. Nonetheless, in agreement with \cite{Nott15}, we found
within the [NaCl] range studied that the entropic contribution 
$\Delta S$ to the FH $\chi$ parameter is always negative, meaning 
that entropy favors phase separation.

The full FH phase diagrams are provided in Fig.~\ref{fig:Bd-FH_and_OV-DH}a.
Three features are notable. 
First, the FH phase boundaries always concave downward, whereas
RPA phase boundaries are S-shaped 
with inflection points, i.e., part of it convex downward
(Fig.~\ref{fig:Ddx4_N1-N1CS_compare}). 
Second, FH predicts a salt-independent 
critical concentration, $\phicr \approx 1/\sqrt{N}$. 
Third, the FH phase boundaries of different salt concentrations cross at 
$T\approx 100^{\circ}$C, exhibiting opposite tendency above and below 
this temperature. These features, which are testable by experiment
in principle, are not presented in the corresponding RPA phase boundaries,
as will be discussed in Sec.~\ref{sec:augRPA} below.

\subsection{The Overbeek-Voorn/Debye-H\"uckel (OV/DH) model}

The OV/DH model is based on the linearized Poisson-Boltzmann 
equation, which self-consistently calculates the screening effect 
of monomeric ions~\cite{Overbeek57,DHmodel}. 
To apply DH to polyampholytes such as Ddx4, we count all charged 
residues as independent monomeric ions and calculate the electrostatic 
energy as
\begin{equation}
f_{\rm DH}  = 
-\frac{1}{12\pi}\left( \frac{4\pi [ (\phi_m/N)(\sum_i|\sigma_i| 
+ |\sum_i\sigma_i|) + 2\phi_s]}{T^*}  \right)^{3/2} \; .
	\label{eq:fDH}
\end{equation}
In salt-free solution, the OV/DH model entails an interaction 
$\propto\phi_m^{3/2}$, instead of $\propto\phi_m^2$ in FH.
as noted recently in the context of IDP phase separation~\cite{Brangwynne15}.
Because $\partial^2\phi_m^{3/2}/\partial\phi_m^2$ $\sim$ $\phi_m^{-1/2}$,
it predicts a $\phi_m$ dependence that is much sharper than the 
corresponding constant term in FH. This feature reflects the long-range
nature of electrostatic interactions and allows 
phase separation to occur under much more dilute concentrations. 

To compare OV/DH model predictions
with experiment, values for the link length scale $b$ and 
dielectric constant $\epsilon_r$ are needed in Eq.~(\ref{eq:Teq}) 
to convert the reduced temperature $T^*$ in the model to actual 
temperature. As in our previous RPA work,
we take $b$ to be the C$\alpha$-C$\alpha$ virtual bond length 3.8\AA, and 
let $\epsilon_r$ be a fitting parameter~\cite{Lin16}, as $\epsilon_r$ 
of an aqueous protein solution can vary widely between $\approx$ 2 
and 80~\cite{Pitera01, Warshel06, Leach}. A dielectric constant 
$\epsilon_r =59.5$ resulted from fitting with experimental data
(Fig.~\ref{fig:Bd-FH_and_OV-DH}b, inset).
The corresponding DH phase diagrams are provided in 
Fig.~\ref{fig:Bd-FH_and_OV-DH}b. In comparison with 
the FH phase diagrams in Figs.~\ref{fig:Bd-FH_and_OV-DH}a, 
OV/DH predicts lower $\Tscr$ and salt-dependent $\phicr$ instead
of the salt-independent $\phicr$ in FH. Unlike FH, the phase boundaries 
of OV/DH model do not cross each other.

\begin{figure*}[t]
\includegraphics[width=\textwidth]{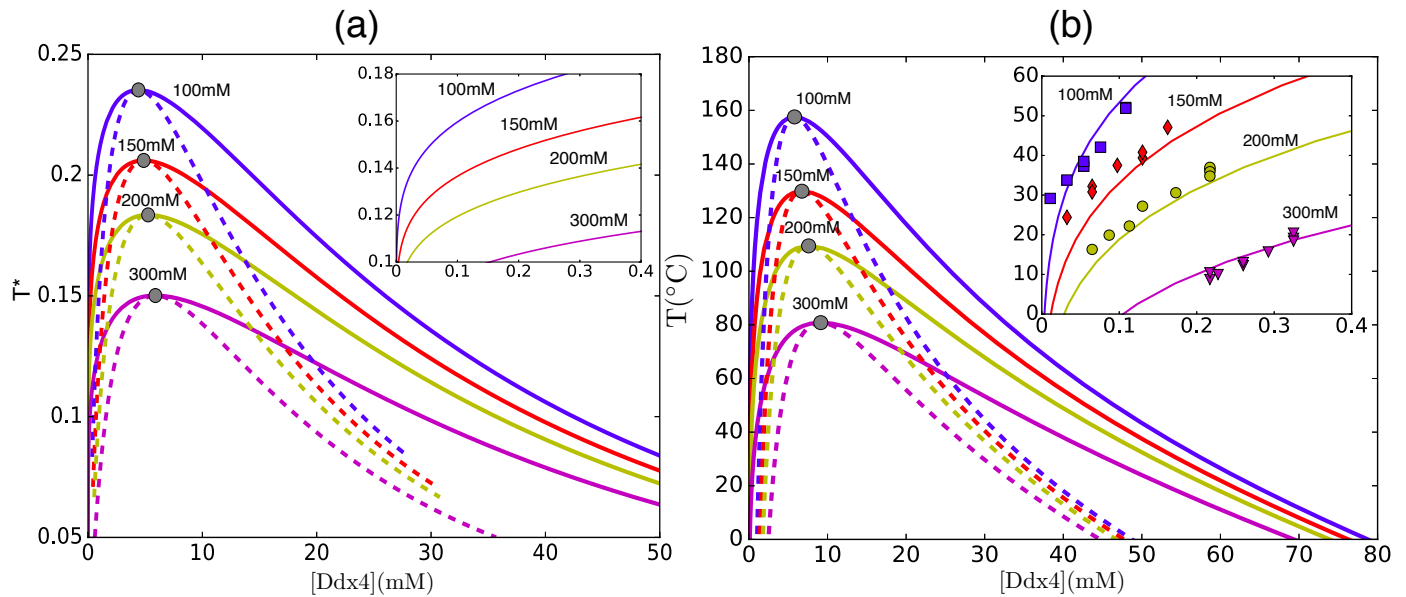}  
\caption{ (a) Theoretical phase diagrams of Ddx4$^\text{N1}$ with 
different [NaCl]'s, computed by (a) Eq.~(\ref{eq:fel_ori}), 
the RPA-only theory, and (b) Eq.~(\ref{eq:fel_ori}) with 
Eq.~(\ref{eq:fFH_in_RPAFH}), the RPA+FH theory. 
Binodal and spinodal boundaries are shown, respectively, as solid and
dashed lines. Critical points are marked by gray circles.
The two insets highlight the phase behavior for small Ddx4 concentration
for which experimental data are availabe~\cite{Nott15} (shown as
symbols with the same color code in the inset of (b)).
Data in the insets were provided previously \cite{Lin16} and are included 
here for completeness. }
	\label{fig:RPA_and_RPA_FH}
\end{figure*}

\subsection{An augmented RPA model}
\label{sec:augRPA}

We recently applied RPA and an augmented RPA+FH model to the 
salt-dependent phase behavior of Ddx4$^{\rm N1}$~\cite{Lin16}, 
focusing only on the low [Ddx4] regime for which experimental data 
are available (Fig.~\ref{fig:RPA_and_RPA_FH}, insets). Here we extend 
that theoretical study to explore essentially the entire binodal and 
spindoal boundaries of Ddx4$^{\rm N1}$ at four [NaCl]'s
(Fig.~\ref{fig:RPA_and_RPA_FH}), and compare their properties
with the FH and OV/DH  theories discussed above.

Our RPA model for Coulomb interaction may be viewed 
as a one-step improvement from the DH model. As described in 
Sec.~\ref{sec:seq-specified_RPA}, RPA treats polymers 
as Gaussian chains without excluded volume. It allows a direct
plug-in of any given charge pattern onto the chain sequence.
Different charge patterns distribute their 
interactions over the Gaussian-chain conformations differently.
Therefore, despite the approximations involved in RPA \cite{Borue88,Borue90},
it does provide an account of sequence dependence that is absent in
the FH and OV/DH approaches. In particular, RPA allows us to 
rationalize and gain insights into the different behaviors of 
Ddx4$^{\rm N1}$ and Ddx4$^{\rm N1}$CS (Fig.~\ref{fig:Ddx4_N1-N1CS_compare}) 
\cite{Lin16}, which is not possible, at least not in any conceptually
clear manner, for FH and OV/DH. In the special case when the polymer length
in RPA is equal to one, RPA is reduced to DH.

Similar to the OV/DH phase boundaries in Fig.~\ref{fig:Bd-FH_and_OV-DH}b, 
the RPA phase boundaries of different [NaCl]'s calculated using
Eq.~(\ref{eq:fel_ori}) do not cross each other 
(Fig.~\ref{fig:RPA_and_RPA_FH}a). Notably, the RPA phase boundaries 
are S-shaped, unlike the inverted U-shaped curves in FH
(Fig.\ref{fig:Bd-FH_and_OV-DH}a). For the OV/DH model, the spinodal
boundaries are also clearly S-shaped; some of the binodal boundaries 
are also slightly S-shaped (Fig.\ref{fig:Bd-FH_and_OV-DH}b), 
but not as pronounced as for the RPA and RPA+FH models.
S-shaped phase boundaries in RPA models have
also been reported in polyelectrolyte systems, and have been
proposed in those cases as a consequence of formation of a polymer 
network in the solution~\cite{Mahdi00}. 
In the context of IDP phase separation, S-shaped phase boundaries
may help rationalize the possibility of having a relatively low IDP 
concentration even in the phase-separated IDP-rich phase.
So far our RPA+FH theory has only been tested within
a very narrow low concentration range for which experimental data
are available (inset of Fig.~\ref{fig:RPA_and_RPA_FH}b).
More extensive experimental data and numerical simulations will be 
needed to test the theoretically predicted phase boundary to the 
right of the critical point.

As we have pointed out in \cite{Lin16}, while RPA rationalizes
the trend of salt dependence (Fig.~\ref{fig:RPA_and_RPA_FH}a, inset),
RPA by itself does not account for the effect quantitatively.
For this reason and the physical consideration that short-range 
aromatic cation-$\pi$ and $\pi$-$\pi$ interactions~\cite{Ma97, Meyer03} 
are important for IDP interactions
in general~\cite{Chen15, Song13,Nott15, Schmidt15, Brangwynne15, Vernon16}
and play a significant role in Ddx4 phase separation
in particular~\cite{Nott15}, we constructed an RPA+FH model by augmenting
the $f_{el}$ in Eq.~(\ref{eq:fel_ori})
with the following salt-independent FH interaction term,
\begin{equation}
f_{\rm FH} = \chi\phi_m(1-\phi_m) = \left(\frac{\varepsilon_h}{T^*}+ 
\varepsilon_s\right)\phi_m(1-\phi_m) \; ,
	\label{eq:fFH_in_RPAFH}
\end{equation}
such that the $f_{\rm int}$ in the total free energy 
$f$ in Eq.~(\ref{first_eq}) becomes
\begin{equation}
f_{\rm int} = f_{\rm el} + f_{\rm FH} \; .
\end{equation}
Here $\varepsilon_h$ and $\varepsilon_s$ are the enthalpic and entropic 
contributions, respectively, to the FH interaction parameter $\chi$. 
By treating $\varepsilon_h$ and $\varepsilon_s$ as global 
fitting parameters, a reasonably good fit is achieved with experimental 
results for all four available [NaCl] values~\cite{Nott15} 
(Fig.~\ref{fig:RPA_and_RPA_FH}b, inset)
by using a value of $\epsilon_r=29.5$ which is physically
plausible for a low-concentration IDP solution (see Sec.~\ref{epsilon_sec}
below).
The overall shapes of the phase boundaries 
remain largely unchanged vis-\`a-vis those predicted by RPA 
(cf. Fig.~\ref{fig:RPA_and_RPA_FH}a and b). Similar to the OV/DH
model (Fig.~\ref{fig:Bd-FH_and_OV-DH}b), the critical concentration 
increases with increasing salt 
in both the RPA and RPA+FH models, in contrast to the salt-independet
critical concentration predicted by FH (Fig.~\ref{fig:Bd-FH_and_OV-DH}a).

As we have discussed previously, 
the fitted $\varepsilon_h = 0.15$ and $\varepsilon_s = -0.3$ 
used to produce Fig.~\ref{fig:RPA_and_RPA_FH}b may be rationalized
to an extent by considering the strength of cation-$\pi$ interactions
as well as the expected loss of sidechain conformational entropy
upon formation of such contacts \cite{Lin16}. Of course, given that RPA
is an approximate approach, $f_{\rm FH}$ may also serve to
correct inaccuracies in the pure-RPA theory. Finally, it is noteworthy
that our fitted entropic free energy $\varepsilon_s = -0.3$ is negative, 
which means that the entropy in the FH $\chi$ parameter is positive
(because of the $-T$ factor needed to convert entropy to free energy).
Hence the entropy in our FH $\chi$ disfavors phase separation whereas
that in the FH model in Sec.~\ref{FH_subsection} favors phase separation. 
Nonetheless, the two results are not necessarily inconsistent because
there are additional contributions to conformational entropy in
the RPA $\fel$, namely the $\hat{G}_k$ in Eq.~(\ref{eq:fel_ori}) that 
accounts for conformational freedom of Gaussian chains.

\section{Possible cooperativity-enhancing effects of
concentration dependent relative permittivity}
\label{epsilon_sec}

In all the theories considered above, the relative permittivity, or 
dielectric constant $\epsilon_r$, is considered to be a constant 
throughout the solution irrespective of whether the system has phase 
separated. However, because proteins have very different relative
permittivity ($\epsilon_r \approx 2$ -- $4$) from that of bulk 
water ($\epsilon_r\approx 80$) \cite{Leach}, $\epsilon_r$ is
expected to depend on IDP concentration. Although a thorough
examination of this issue is beyond the scope of this work, here
we take a first step to explore how a concentration-dependent
$\epsilon_r(\phi_m)$ might affect phase properties.

\begin{figure}[t]
\includegraphics[width=\columnwidth]{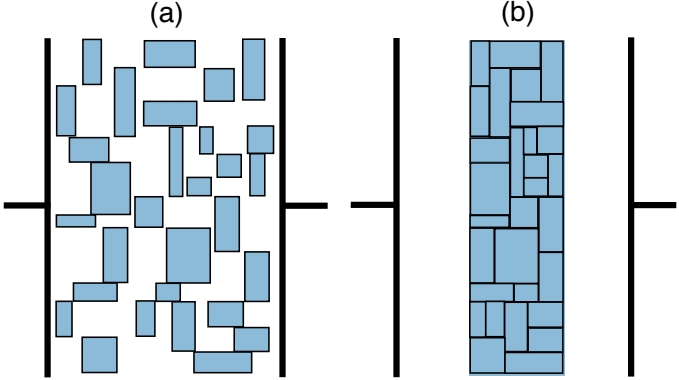}  
\caption{Estimating effective dielectric constants for IDP-water
mixtures by consideration of a simple parallel plate capacitor. Here the 
IDP material is depicted by the shaded areas and water is represented
by the white space. The electric capacitance of configurations (a) and
(b) are identical irrespective of the dispersion of the IDP material
as long as the total width of IDP material is constant over the
entire area of the parallel plates (i.e., always constituting $\phi_m$
of the distance between the plates). This consideration
allows the effective dielectric constant of the mixture in (a)
to be determined by standard application of Gauss' law to 
(b) under the assumption that the plates are
sufficiently large such that edge effects are negligible,
as presented in elementary physics textbooks (see, e.g.,~\cite{GaussLaw}).}
	\label{fig:charged_slabs}
\end{figure}

\begin{figure}[t]
\includegraphics[width=\columnwidth]{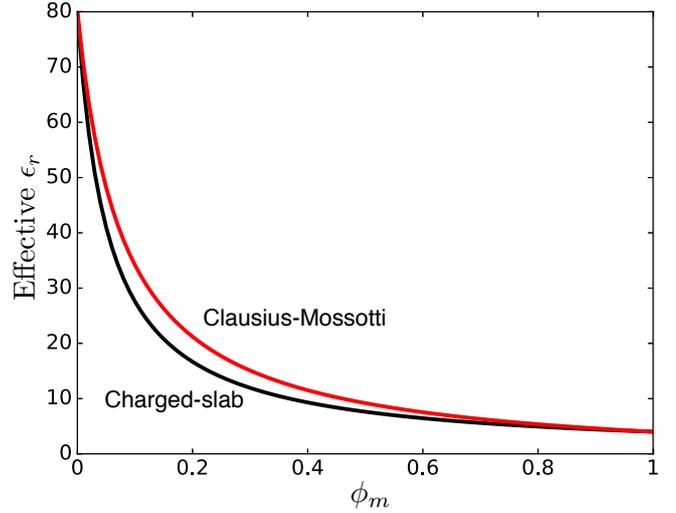}  
\caption{Charged-slab and Clausius-Mossotti models for 
concentration-dependent relative permittivity. 
The $\epsilon_r(\phi_m)$ plotted here are the 
$\epsilon_{\rm eff}$ functions given in Eqs.~(\ref{eq:e_eff_simple})
and (\ref{eq:e_eff_CM_eq}) for $\epsilon_p = 4$ and $\epsilon_w = 80$. 
The two models predict similar $\phi_m$ dependence.}
	\label{fig:dic_two_models}
\end{figure}

One way to estimate the effective $\epsilon_r$ of a mixture of protein
(IDP) and water is to consider a pair of infinitely large charged slabs
(an electric capacitor) with protein and solvent in between. 
For a solution in which protein volume ratio is $\phi_m$ and water 
is $1-\phi_m$, the electric field $E$ between the two slabs is 
given by Gauss' law~\cite{GaussLaw}:
\begin{equation}
\epsilon_0 E = \phi_m \frac{\sigma}{\epsilon_p} +  
(1-\phi_m) \frac{\sigma}{\epsilon_w} \equiv \frac{\sigma}{\epsilon_{\rm eff}},
\end{equation}
where $\sigma$ is the charge density on the slabs and $\epsilon_p, 
\epsilon_w$ are the relative permittivity, respectively, of protein
and water. It follows that the effective relative permittivity is
given by
\begin{equation}
\epsilon_{\rm eff}(\phi_m) = 
\frac{\epsilon_p \epsilon_w}{\phi_m \epsilon_w + (1-\phi_m)\epsilon_p}.
	\label{eq:e_eff_simple}
\end{equation}
We expect this expression to be quite general because for a 
given $\phi_m$, application of Gauss' law leads to the same 
$\epsilon_{\rm eff}$, almost -- albeit not entirely -- irrespective of 
the configuration of the mixture of the two media, 
as is illustrated in Fig.~\ref{fig:charged_slabs}.
An alternative estimate of $\epsilon_{\rm eff}$ is by considering
the dipole moments and molecular polarizabilities of water and 
protein components and apply the Clausius-Mossotti 
equation~\cite{CM_equation}, which leads to
\begin{equation}
\epsilon_{\rm eff}(\phi_m) = \frac{1+2[(1-\phi_m)\gamma_w + 
\phi_m \gamma_p ]}{1-[(1-\phi_m)\gamma_w + \phi_m \gamma_p ]} \; ,
	\label{eq:e_eff_CM_eq}
\end{equation}
where
\begin{equation}
\gamma_w \equiv \frac {\epsilon_w -1}{\epsilon_w + 2} \; ,
\quad \quad
\gamma_p \equiv \frac {\epsilon_p -1}{\epsilon_p + 2} \; ,
\end{equation}
are proportional to the molecular polarizabilities of water
and protein, respectively.
The dependences of $\epsilon_{\rm eff}$ on $\phi_m$ estimated from
the two approaches are very similar (Fig.~\ref{fig:dic_two_models}),
buttressing our expectation that the predicted trend is robust.
Interestingly, $\epsilon_{\rm eff}$ decreases sharply from that
of bulk water for relative small $\phi_m$. For instance,
$\epsilon_{\rm eff}$ decreases from 80 to $\approx 20$
for $\phi_m\approx 20\%$. This effect is significant because
according to our RPA+FH model (Fig.~\ref{fig:RPA_and_RPA_FH}b), 
$\phi_m$ for the
condensed phase of Ddx4 at 60$^\circ$C and [NaCl]= 100mM 
is $\approx 40\times 10^{-3}\times 236/55.5 = 17\%$,
suggesting that an IDP-concentration-dependent relative permittivity 
should contribute to a more complete physical account.

We now incorporate this effect into our RPA theory by replacing
$l_B$ in Eq.~(\ref{eq:Uk}) by
$l_B = e^2/(4\pi\epsilon_0\epsilon_r(\phi_m) k_{\rm B} T)$,
where $\epsilon_r(\phi_m)$ is given by Fig.~\ref{fig:dic_two_models}
and, in place of the reduced temperature $T^*$ in Eq.~(\ref{eq:Teq}),
define another reduced temperature $T_0^*$ in terms of vacuum permittivity
$\epsilon_0$, viz.,
\begin{equation}
T_0^* \equiv {4\pi\epsilon_0 k_{\rm B} Tb}/{e^2} \; .
	\label{eq:T0eq}
\end{equation}
In addition, because ${\cal G}$ in Eq.~(\ref{eq:fel_dimless}) is no
longer linear in $\phi_m$ because of $\epsilon_r(\phi_m)$, it cannot
be used as the last term in the curly brackets of Eq.~(\ref{eq:fel_dimless}) 
for the sole purpose of subtracting out unphysical ultraviolet divergences
because including a ${\cal G}$ not linear in $\phi_m$ would
affect phase properties. Consequently, for the case with
$\epsilon_r(\phi_m)$, we must derive the
dimensionless $\fel$ directly from Eq.~(\ref{eq:fel_ori}), yielding 
\begin{equation}
\fel = \int \frac{d\kr\kr^2}{4\pi^2} \left\{\frac{1}{\eta}\ln\left[ 
1 + \eta {\cal G}_1(\kr)\right] - {\cal G}_2(\kr) \right\} \; ,
	\label{eq:fel_dimless_dic}
\end{equation}
where, instead of ${\cal G}$ in Eq.~(\ref{eq:G_cal}), we now have
${\cal G}_1$ for the determinant and ${\cal G}_2$ for the trace,
\begin{subequations}
\begin{align}
\!\!{\cal G}_1(\kr) & = \frac{4\pi}{\kr^2[1+\kr^2] T_0^* \epsilon_r(\phi_m)}
         \left( 2\phi_s \!+\! \phi_c \!+\! \frac{\phi_m}{N}  
         	\langle \sigma | \GM(\kr) | \sigma \rangle \right), \\
\!\!{\cal G}_2(\kr) & = \frac{4\pi}{\kr^2[1+\kr^2] T_0^* \epsilon_r(\phi_m)}
	\left(2\phi_s \!+\! \phi_c \!+\! \frac{\phi_m}{N} \sum_i|\sigma_i|  
\right) \; .  
\end{align}
\end{subequations}

\begin{figure}[t]
\includegraphics[width=\columnwidth]{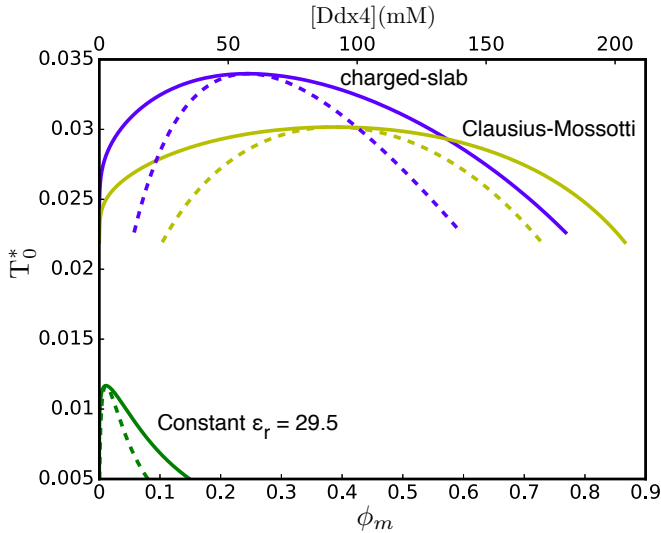}  
\caption{The salt-free phase diagrams of the two $\epsilon_r(\phi_m)$ 
models in Eqs.~(\ref{eq:e_eff_simple}) and (\ref{eq:e_eff_CM_eq})
for Ddx4$^{\rm N1}$. 
Solid and dashed lines are, respectively, binodal and spinodal boundaries. 
IDP and water relative permittivity of $\epsilon_p = 4$ and $\epsilon_w = 80$,
respectively, are used to calculate $\epsilon_r(\phi_m)$ in both the 
charged-slab and Clausius-Mossotti models.}
	\label{fig:three-model_compare}
\end{figure}

Fig.~\ref{fig:three-model_compare} compares
the salt-free phase diagrams predicted by this new theory with
that obtained using our original RPA theory with $\phi_m$-independent
$\epsilon_r = 29.5$. The result indicates that IDP-concentration-dependent
$\epsilon_r$ can lead to dramatic shifts of phase boundaries.
Relative to the constant-$\epsilon_r$ model, both $\epsilon_r(\phi_m)$ 
models have a significantly higher critical temperature and thus a
much higher tendency to phase separate. At a qualitative level,
the phase-separation enhancing effect of $\epsilon_r(\phi_m)$ is
not difficult to understand. As $\epsilon_r(\phi_m)$ decreases
with increasing $\phi_m$ (Fig.~\ref{fig:dic_two_models}),
electrostatic interactions favoring formation of a concentrated
phase become even stronger with increasing $\phi_m$ than if 
$\epsilon_r$ is a constant. This increases the favorability of
the concentrated phase, and amounts to a form of many-body 
electrostatic mechanism that enhances phase transition cooperativity, 
playing a role similar in effect, though not necessarily in origin, to 
various many-body mechanisms in cooperative protein 
folding~\cite{artem08,chan2011}. This exploratory study has thus
demonstrated the potential importance of $\epsilon_r(\phi_m)$-related
effects. However, many aspects of these effects, such as possible
distance dependence of relative permittivity \cite{Leach,Freed05}, 
remain to be investigated. When these issues are addressed in the
future, attention should also be paid to the possibility that the 
tendency for cooperative condensation may be overestimated by using 
a mean-field concentration $\phi_m$~\cite{Wang14}, because such
cooperativity is likely to be attentuated when constraints from 
chain connectivity are fully taken into account.

\section{Conclusion}

In summary, we have presented in some detail a general analytical 
RPA theory to account for sequence-specific electrostatic effects on 
polyampholyte phase separation, and demonstrated its utility
by applying our theory to Ddx4. Using Ddx4 as an example, we have 
compared the predicted phase behaviors of RPA and an augmented RPA 
model with those predicted by FH or OV/DH theories, identifying
several salient differences. We have also outlined IDP 
concentration-dependent relative permittivity as a potentially
productive direction for theoretical progress.
The predicted phase properties by all of the theories documented
here are testable by future experiments. 

In the study of IDP phase separation, because direct simulations 
of multiple-polymer systems are computationally costly, tractable 
analytical theories are extremely useful for conceptual development, 
physical insights, and generating new ideas for experiment.
However, direct computer simulations of phase separation
are indispensable, because they are necessary for checking the 
validity of the assumptions and approximations in analytical 
models~\cite{Orkoulas03}.
RPA assumes uniform densities of counterions and salts. As such
it is not expected to be adequate in situations when counterion
condensation is significant, such as in the recently observed 
sequence-dependent coacervation of the Nephrin intracellular 
domain~\cite{RosenPappu2016}. Computer simulations and further 
development of analytical theory, such as the incorporation of
formulations that are capable of accounting for effective correlations 
beyond second order in density fluctuations \cite{Ghosh15,Muthukumar96}, 
would be needed in future attempts to address such behaviors. 

We deem it fitting to dedicate this paper to the celebration of
Professor Vojko Vlachy's seventieth birthday. Among many
of his scientific achievements, Vojko has made seminal 
contributions to the theory of liquid-liquid phase separation 
of folded proteins~\cite{Vlachy93} and electrostatic 
effects in molecular liquids~\cite{Vlachy99}. In view of
the recent discovery of polyampholytic IDP phase separation as a 
major physical basis for membraneless organelles, his pioneering
work will no doubt be instructive for future theoretical 
endeavors to decipher these novel, fascinating biological 
phenomena. By the same token, experimental techniques that have been 
applied to study phase separation of folded proteins, such as applications 
of hydrostatic pressure~\cite{Royer2011,Silva2014,Dias2014} to investigate
phase separation of folded lysozyme~\cite{Schroer11,Moeller14}, may
also be brought to bear to enrich our biophysical understanding of 
sequence-dependent IDP phase separation.

\section*{Acknowledgements}

We thank Andy Baldwin, Kingshuk Ghosh, Lewis Kay, and Rohit Pappu for helpful 
discussions, and Timothy Nott for the spinodal decomposition images in Fig.7.
H.S.C. wishes to take this opportunity to express his gratitude to 
Ken Dill for introducing him to theoretical studies of 
biopolymers thirty years ago when they first met in late 1986.
This work was supported by a Canadian Cancer Society Research 
Institute grant to J.D.F.-K. and H.S.C. and a Canadian Institutes of Health 
Research grant to H.S.C. We are grateful to SciNet of Compute Canada for
their generous allotment of computational resources.

\appendix

\section{The general binodal condition for an arbitrary number of 
solute components} \label{app:binodal_general}

For a solution with $m$ types of molecules labeled by $i$ with length
(number of monomers) $N_i$ and monomer size $r_i$, and the number 
of type-$i$ molecules being $n_i$, 
the total number of lattice sites
$M = \sum_{k=1}^m r_k n_k N_k$ and the free energy per site
may be expressed as a function $f(\{ \phi_k \})$ of the $m-1$ independent
densities $\phi_i$ defined in Eqs.~(\ref{eq:phi_i}) 
and (\ref{eq:phi_normalize}). Note that there are only $m-1$ independent
$\phi_i$'s ($i=1,2,\dots,m-1$) because of Eq.~(\ref{incompressibility_eq}). 
The molecule labels are chosen such that $i=m$ for the
water solvent.

The chemical potential $\mu_i$ of type-$i$ molecules in units of
$k_BT$ is the derivative of the free energy $F(\{\phi_k \}) =$
$Mf(\{ \phi_k \})$ in Eq.~(\ref{first_eq}) with respect to
the number $n_i$ of type-$i$ molecules~\cite{Voorn56a, Grosberg94, chandill94}:
\begin{equation}
\begin{aligned}
\frac{\mu_i}{k_B T} = 
& \frac{1}{k_B T}\frac{\partial F(\{ \phi_k \} )}{\partial n_i}  
=  \frac{\partial Mf(\{ \phi_k \} ) }{\partial n_i} \\
= & \frac{\partial \sum_k r_k n_k N_k }{\partial n_i} f + M \frac{\partial f(\{ \phi_k \})}{\partial n_i} \\
= & r_i N_i f + M \sum_{j=1}^{m-1} \frac{\partial f}{\partial \phi_j} \frac{d \phi_j}{d n_i} \\
= & r_i N_i f + M \sum_{j=1}^{m-1} \frac{\partial f}{\partial \phi_j} \left( \delta_{ij}\frac{N_j}{M} - \frac{ n_j N_j r_i N_i}{M^2}  \right) \\
= & N_i \left( r_i f + t_i f_i' - r_i\sum_{j=1}^{m-1} \phi_j f_j'   \right),
\end{aligned}
	\label{eq:mu_i_general}
\end{equation}
where $\delta_{ij}$ is the Kronecker symbol, $\delta_{ij}=1$ for $i=j$;
$\delta_{ij}=0$ for $i\ne j$, the symbol $t_i = 1$ for $1 \leq i \leq m-1$ 
and $0$ for $i = m$, and $f_i' = \partial f/\partial \phi_i$ is the 
first-order derivative of $f$ with respect to $\phi_i$. When molecules
of type $i$ are carrying charges with valance $z_i$ 
and the Coulomb potential is $\psi^{\alpha}$ in phase $\alpha$, 
the electrochemical potential
\begin{equation}
\eta_i^{\alpha} = \mu_i^{\alpha} + z_i e \psi^{\alpha}
\end{equation}
is the relevant variable for determining the binodal boundary~\cite{Voorn56a}.
When the system is demixed to $n_{\rm ph}$ different phases in equilibrium, 
the chemical, or more generally the electrochemical potentials in all 
phases have to be identical. In other words,
\begin{equation}
\mu_i^{(1)} + z_i e \psi^{(1)} = \mu_i^{(2)} + z_i e \psi^{(2)} = ... 
= \mu_i^{(n_{\rm ph})} + z_i e \psi^{(n_{\rm ph})}
	\label{eq:mu_i_equal}
\end{equation}
for all $1 \leq i \leq m$ molecule types. This condition provides
$m(n_{\rm ph}\!-\!1)$ equations for a total of $n_{\rm ph}(m\!-\!1)$ 
densities/concentrations $\phi_i^\alpha$ of all molecule types in
all different phases. 
The maximum number of separated phases $n_{\rm ph,max}$ for a
set of environmental conditions such as temperature, pressure, etc. 
is then given by $m(n_{\rm ph,max}\!-\!1) = n_{\rm ph,max}(m\!-\!1)$, 
implying $n_{\rm ph,max}=m$ which, as expected, satisfies Gibbs' 
phase rule~\cite{Atkins09}. For $n_{\rm ph} < m$, multiple 
sets of $\{ \phi_m^\alpha\}$ are allowed, with the sets forming 
an $(m\!-\!n_{\rm ph})$-dimensional surface in the $(m\!-\!1)$-dimensional 
$\phi$-space. 
 
We now apply the above general formulation to our polyampholyte system 
with at most $n_{\rm ph}=2$ separated phases 
and five different molecule types: 
polyampholytes with length $N$, size $r_m$, and concentration (density) 
$\phi_m$; counterions of size $r_c$, concentration $\phi_c$; positive 
and negative salt ions with size $r_s$, concentration $\phi_{s\pm} = \phi_s$;
and water of size $r_w$ and $\phi_w =$ 
$(1-r_m\phi_m-r_c\phi_c-2r_s\phi_s)/r_w$ [Eq.~(\ref{incompressibility_eq})]. 
Instead of considering only two concentrations $\phi_m$ and $\phi_s$ as 
in our discussion of the spinodal condition for our previous 
Ddx4 model \cite{Lin16} in Eq.~(\ref{eq:f_phim+phis}), 
here we consider all five concentrations separately for the more general 
case in which $z_i$'s can be different and the system as a whole may 
not be electrically neutral.
Substituting the five molecule types for the $\phi_i$'s in 
Eq.~(\ref{eq:mu_i_general}) and subtracting the equation for 
$\mu_w$ from other $\eta_i$ equations in Eq.~(\ref{eq:mu_i_equal}), 
we arrive at the following equations for binodal (coexistence) 
conditions between phases $\alpha$ and $\beta$ ($\alpha,\beta=1,2$),
in obvious notation:
\begin{subequations}
\begin{align}
f_m'^\alpha + \sigma e \psi^\alpha 
	& = f_m'^{\beta} + \sigma e \psi^{\beta}, \label{eq:df_m} \\ 
f_c'^\alpha -{\rm sign}(\sigma) e \psi^\alpha  
	& = f_c'^{\beta} - {\rm sign}(\sigma)e \psi^{\beta}, \label{eq:df_c} \\
f_{s+}'^\alpha + e \psi^\alpha 
	& = f_{s+}'^{\beta} +  e \psi^{\beta}, \label{eq:df_sp} \\
f_{s-}'^\alpha - e \psi^\alpha 
	& = f_{s-}'^{\beta} -  e \psi^{\beta}, \label{eq:df_sm} \\
f^{\alpha} - \phi_m^{\alpha} f_m'^\alpha - \phi_c^{\alpha} f_c'^\alpha - 
	& \phi_{s+}^{\alpha} f_{s+}'^\alpha - \phi_{s-}^{\alpha} f_{s-}'^\alpha \nonumber\\ 
= f^{\beta} - \phi_m^{\beta} f_m'^\beta - \phi_c^{\beta} 
	& f_c'^\beta - \phi_{s+}^{\beta} f_{s+}'^\beta- \phi_{s-}^{\beta} f_{s-}'^\beta, \label{eq:mu_w}
\end{align}
	\label{eq:binodal_cond_ori}%
\end{subequations}
where $\sigma = (1/N)\sum_{i=1}^N \sigma_i$ is the overall average 
charge density of the polyampholyte. Thus, when the neutrality condition
is applied, $\phi_c = |\sigma|\phi_m$. The above five conditions can 
be further simplified for our RPA model by evaluating each of the $f'$s 
using the expressions for $s$ and $f_{\rm el}$ in Eqs.~(\ref{eq:entropy}) 
and (\ref{eq:fel_dimless}), resulting in a free energy that depends 
only on two concentrations, $f = f(\phi_m, \phi_s)$. The corresponding
binodal conditions are merged into three equations:
\begin{subequations}
\begin{align}
& \text{Eq.~(\ref{eq:df_m})} + \sigma \times \text{Eq.~(\ref{eq:df_c})} : \nonumber \\
& \qquad \qquad \left.\frac{\partial f(\phi_m, \phi_s)}{\partial \phi_m}\right|_\alpha = \left.\frac{\partial f(\phi_m, \phi_s)}{\partial \phi_m}\right|_\beta, \label{eq:dfm}\\
& \text{Eq.~(\ref{eq:df_sp})} + \text{Eq.~(\ref{eq:df_sm})} : \nonumber \\
& \qquad\qquad \left.\frac{\partial f(\phi_m, \phi_s)}{\partial \phi_s}\right|_\alpha = \left.\frac{\partial f(\phi_m, \phi_s)}{\partial \phi_s}\right|_\beta, \label{eq:dfs} \\
& \text{Eq.~(\ref{eq:mu_w})} :  \nonumber \\
& \quad f^\alpha 
- \phi_m^\alpha  \left.\frac{\partial f(\phi_m, \phi_s)}{\partial \phi_m}\right|_\alpha 
- \phi_s^\alpha \left.\frac{\partial f(\phi_m, \phi_s)}{\partial \phi_s}\right|_\alpha \nonumber \\
& \quad = f^\beta 
- \phi_m^\beta \left.\frac{\partial f(\phi_m, \phi_s)}{\partial \phi_m}\right|_\beta 
- \phi_s^\beta \left.\frac{\partial f(\phi_m, \phi_s)}{\partial \phi_s}\right|_\beta \label{eq:muw}.
\end{align}
	\label{eq:binodal_cond}%
\end{subequations}
Note that all linear terms of $\phi_m$ and $\phi_s$ in $(\phi_m, \phi_i)$ 
have no effect in Eq.~(\ref{eq:binodal_cond}) because they only add
an equal constant to both sides of Eqs.~(\ref{eq:dfm}) and (\ref{eq:dfs})
and they cancel out by themselves on each side of  Eq.~(\ref{eq:muw}). 

The conditions in Eq.~(\ref{eq:binodal_cond}) are equations for
the common tangent of free energy function $f(\phi_m, \phi_s)$ at
two points $\alpha = (\phi_m^\alpha, \phi_s^\alpha)$ and 
$\beta = (\phi_m^\beta, \phi_s^\beta)$. Because there are only 
three equations in Eq.~(\ref{eq:binodal_cond}) for the four 
variables $\phi_m^\alpha$, $\phi_m^\beta$, $\phi_s^\alpha$, 
and $\phi_s^\beta$, the common-tangent pair $(\alpha, \beta)$ 
is not unique determined. A given temperature $T^*$ is consistent
with a series of $(\alpha, \beta)$. These common-tangent sets constitute 
a closed contour on the $\phi_m$--$\phi_s$ surface as the 
binodal boundary~\cite{Voorn56c}. A system with a set of original 
bulk concentrations $O = (\phi_m^0, \phi_s^0)$ inside the contour 
will phase separate to two regions with concentrations 
$(\phi_m^\alpha, \phi_s^\alpha)$, $(\phi_m^\beta, \phi_s^\beta)$ such that 
$O, \alpha, \beta$ are on the same common tangent. In other words,
a fourth equation for solving a unique set of $(\alpha, \beta)$ 
is provided by the initial concentrations.

\section{Critical points of strictly alternating polyampholytes in
the $N\to\infty$ limit}~\label{app:alter_N_infty}

In the limit of $N\to\infty$, the parameters $A$ and $B$ in Eq.~(\ref{eq:fel_infty_phim3}) can be expressed as
\begin{subequations}
\begin{align}
A = & A_r(\sigma) \sigma^{3/2} T^{*-2}, \\
B = & B_r(\sigma) \sigma^{3/2} T^{*-3}, 
\end{align}
\end{subequations}
where $A_r$ and $B_r$ are functions of $\sigma$ via the following
integrals,
\begin{subequations}
\begin{align}
A_r(\sigma) = & \sqrt{\frac{8}{3}}
	\int_0^\infty \!\! \frac{dk'}{k'^2(1+6\sigma k'^2)^2}\left( \frac{1-e^{-k'^2}}{1+e^{-k'^2}} \right)^2 \\
B_r(\sigma) = & \sqrt{\frac{128}{27}} \pi 
	\int_0^\infty \!\!\frac{dk'}{k'^4 (1+6\sigma k'^2)^3}\left( \frac{1-e^{-k'^2}}{1+e^{-k'^2}} \right)^3.
\end{align}
\end{subequations}
As $\sigma \to 0$, $A_r\to1.374$ and $B_r\to1.187$, as has been 
shown in Eqs.~(\ref{eq:chi_infty}) and (\ref{eq:B+for_fel_3rd_div}). For 
general $\sigma \leq 1$, $A_r$ and $B_r$ become much smaller because 
of the short range cutoff $(1 + 6\sigma k'^2)$.

To self-consistently solve the critical point, we calculate the 
derivatives of free energy $f = -s + \fel^\infty$,
\begin{equation}
\left.\frac{\partial^2 f}{\partial \phi_m^2} 
\right|_{\phi_m = \phicr, T^* = \Tscr} = 0 \; \text{\ and\ }
\left.\frac{\partial^3 f}{\partial \phi_m^3} 
\right|_{\phi_m = \phicr, T^* = \Tscr} = 0, 
\end{equation}
with the $\fel^\infty$ in Eq.~(\ref{eq:fel_infty_phim3}), yielding
\begin{subequations}
\begin{align}
& \frac{1}{N \phicr} + \frac{1}{1-\phicr}  
	- A_r\frac{\sigma^{3/2}}{\Tscrp{2}} 
+ B_r\frac{\sigma^{3/2}}{\Tscrp{3}} \phicr = 0  \; ,
		\label{eq:ddf_for_cri} \\
& -\frac{1}{N \phicr^2} + \frac{1}{(1-\phicr)^2} 
	+ B_r\frac{\sigma^{3/2}}{\Tscrp{3}} = 0 \; .
		\label{eq:dddf_for_cri}
\end{align}
\end{subequations}

We first substitute Eq.~(\ref{eq:dddf_for_cri}) into 
Eq.~(\ref{eq:ddf_for_cri}) for the $B_r$ term to obtain
\begin{equation}
A_r \frac{\sigma^{3/2}}{\Tscrp{2}} = \frac{2}{N\phicr} 
+ \frac{1}{1-\phicr} -\frac{\phicr}{(1-\phicr)^2} \; ,
	\label{eq:Ar_in_solve_cri}
\end{equation}
then we simultaneously express $1/\Tscr$ by Eqs.~(\ref{eq:dddf_for_cri}) 
and (\ref{eq:Ar_in_solve_cri}) to arrive at the following
two equalities:
\begin{equation}
\begin{aligned}
\frac{1}{\Tscr} = & \left[ \left( \frac{2}{N \phicr} 
+ \frac{1}{1-\phicr} - \frac{\phicr}{(1-\phicr)^2} \right) 
\frac{1}{\sigma^{3/2} A_r(\sigma)} \right]^{1/2} \\
= & \left[ \left( \frac{1}{N \phicr^2} - \frac{1}{(1-\phicr)^2} \right) 
\frac{1}{\sigma^{3/2} B_r(\sigma)} \right]^{1/3} \; .
\end{aligned}
	\label{eq:strict_alter_cri_S-A}
\end{equation}
Now $\phicr$ can first be solved numerically using the second equality.
Then $\Tscr$ can be solved by substituting the $\phicr$ value into
the first equality. 

\begin{table}[htbp]
\centering
\begin{tabular}{ccccc}
\hline
$\sigma$ & $\phicr$ (s-a)  & $\phicr$ (num) & $\Tscr$ (s-a) & $\Tscr$ (num) \\
\hline
0.025   & $0.00669$ & $0.00909$ & 0.02833 & 0.03260 \\
0.05     & $0.00880$ & 0.01126  & 0.04148 & 0.04601 \\
0.1       & 0.01111   & 0.01361  & 0.05634 & 0.06091 \\
0.2       & 0.01354  & 0.01604  & 0.07136 & 0.07578 \\
\hline
\end{tabular}
	\caption{The critical point parameters $\phicr$ and $\Tscr$ 
	of $N=240$ strictly alternating polyampholytes, 
	calculated by the semi-analytic
	method (s-a) in Eq.~(\ref{eq:strict_alter_cri_S-A}) and the 
	numerical method (num) described in Sec.~\ref{sec:mean-field_phase}.
	The deviation of the semi-analytic results from the numerical results
        increases as $\sigma$ decreases.
	} 
	\label{table:semi_analytic_cri}
\end{table}

We provide in Table~\ref{table:semi_analytic_cri} the critical points 
of $N=240$ strictly alternating polyampholytes calculated semi-analytically
by pursuing numerical solutions to the 
analytic Eq.~(\ref{eq:strict_alter_cri_S-A}) and 
compare them with the numerical results computed by the method 
described in Sec.~\ref{sec:mean-field_phase}. For $\sigma =0.2$, 
the present semi-analytic (s-a) method
provides a reasonable approximation, with 20\% and 6\% 
deviations in $\phicr$ and $\Tscr$, respectively. Both approaches
yielded $\phicr$ values that are much smaller than the $1/\sqrt{240} = 0.06$ 
value one might otherwise expect from FH. Interestingly, however,
their $\Tscr$ are close to the FH prediction of $\Tscr = 0.03$ -- $0.06$ 
for the different $\sigma$ values tested.

\section*{References}

\bibliographystyle{elsarticle-num-names.bst}
\bibliography{j_mol_liq_v0}

\end{document}